\documentclass[prd,floatfix,twocolumn]{revtex4}

\pdfoutput=1

\usepackage{stmaryrd}

\usepackage{comment}
\usepackage{pdfpages}
\usepackage{graphicx} 
\usepackage{dcolumn} 
\usepackage{amsmath}
\usepackage{amssymb}
\usepackage{bm}
\usepackage{times}
\usepackage{ulem}
\usepackage[justification=raggedright]{caption}

\usepackage{boondox-cal}
\usepackage{subfigure}
\usepackage{tabularx}
\usepackage{appendix}
\usepackage{amstext}
\usepackage{array}

\usepackage{bbm}
\usepackage{blkarray}

\definecolor{wine-stain}{rgb}{0.5,0,0} 
\definecolor{bblue}{rgb}{0,0.0,0.5} 

\usepackage[colorlinks=true
,urlcolor=blue
,anchorcolor=blue
,citecolor=blue 
,filecolor=blue
,linkcolor=wine-stain
,menucolor=blue
,pagecolor=blue
,linktocpage=true
,pdfproducer=medialab
,linktoc=all 
]{hyperref}

\usepackage{booktabs}  

\allowdisplaybreaks

\newcommand{\ncmd}{\newcommand}

\ncmd{\tr}[1]{~\mbox{tr}\lt\{ {#1}\rt\}}
\ncmd{\half}{\frac{1}{2}}
\ncmd{\eps}{\epsilon}
\ncmd{\veps}{\varepsilon}
\ncmd{\dgr}{\dagger}
\ncmd{\abs}[1]{\lt\cb{#1}\rt\cb}
\ncmd{\avg}[1]{\lt\lb{#1}\rt\rb}
\ncmd{\sgn}[1]{\mbox{sgn}\lt(#1\rt)}
\ncmd{\kap}{\kappa}
\ncmd{\wtil}[1]{\widetilde{#1}}
\ncmd{\thrfr}{\therefore}
\ncmd{\eq}[1]{Eq. \eqref{#1}}
\ncmd{\fig}[1]{Fig. \ref{#1}}
\ncmd{\ordr}[1]{\mathcal{O}\lt(#1\rt)}
\ncmd{\dsty}{\displaystyle}
\ncmd{\alert}[1]{\color{red}{#1}}
\ncmd{\mc}{\mathcal}
\ncmd{\mbf}[1]{\mathbf{#1}}
\ncmd{\Deriv}[2]{\frac{d{#1}}{d{#2}}}
\ncmd{\ParDeriv}[2]{\frac{\partial{#1}}{\partial{#2}}}
\ncmd{\td}{\tilde} 
\ncmd{\what}{\widehat}

\ncmd{\Gmnrs}{G_{\mu \nu \rho \sigma}}
\ncmd{\hGmnrs}{ \hat G_{\mu \nu \rho \sigma}}
\ncmd{\Pic}{P}
\ncmd{\Ht}{{\bf H}_\perp}

\ncmd{\sg}{\sqrt{g}} 
\ncmd{\invg}{\frac{1}{g}} 
\ncmd{\invsg}{\frac{1}{\sqrt{g}}} 

\ncmd{\gmn}{g_{\mu \nu}} 
\ncmd{\grs}{g_{\rho \sigma}} 
\ncmd{\hgmn}{\hat g_{\mu \nu}} 
\ncmd{\hgrs}{\hat g_{\rho \sigma}} 
\ncmd{\hg}{\hat g} 
\ncmd{\hsg}{\sqrt{\hat g}} 
\ncmd{\hinvg}{\frac{1}{\hat g}} 
\ncmd{\hinvsg}{\frac{1}{\sqrt{\hat g}}} 

\ncmd{\BL}{\left\llbracket}
\ncmd{\BR}{\right\rrbracket}

\ncmd{\hPi}{\hat \Pi} 
\ncmd{\hPimn}{\hat \Pi^{\mu \nu}} 
\ncmd{\hPirs}{\hat \Pi^{\rho \sigma}} 

\ncmd{\Pimn}{\Pi^{\mu \nu}} 
\ncmd{\Pirs}{\Pi^{\rho \sigma}} 

\ncmd{\Pmn}{P^{\mu \nu}} 
\ncmd{\Prs}{P^{\rho \sigma}}

\ncmd{\lp}{a} 
\ncmd{\da}{\varphi} 
\ncmd{\bgmn}{\bar g_{\mu \nu}}
\ncmd{\bgrs}{\bar g_{\rho \sigma}}
\ncmd{\bgimn}{\bar g^{\mu \nu}}
\ncmd{\bgirs}{\bar g^{\rho \sigma}}
\ncmd{\bgimr}{\bar g^{\mu \rho}}
\ncmd{\bgins}{\bar g^{\nu \sigma}}

\ncmd{\EL}{{\cal N}}
\ncmd{\emtwodba}{e^{-2d \bar \alpha} }
\ncmd{\etwodba}{e^{2d \bar \alpha} }
\ncmd{\efourdba}{e^{4d \bar \alpha} }
\ncmd{\ehalfdba}{e^{\frac{d}{2} \bar \alpha} }
\ncmd{\emdba}{e^{-d \bar \alpha} }
\ncmd{\edba}{e^{d \bar \alpha} }
\ncmd{\emba}{e^{- \bar \alpha} }
\ncmd{\eba}{e^{ \bar \alpha} }
\ncmd{\cM}{{\bf M}}

\ncmd{\xtd}{e^{-2d \bar \alpha} }
\ncmd{\xmtd}{e^{2d \bar \alpha} }
\ncmd{\xmfd}{e^{4d \bar \alpha} }
\ncmd{\xmhd}{e^{\frac{d}{2} \bar \alpha} }
\ncmd{\xd}{e^{-d \bar \alpha} }
\ncmd{\xmd}{e^{d \bar \alpha} }
\ncmd{\xnot}{e^{- \bar \alpha} }
\ncmd{\xmnot}{e^{ \bar \alpha} }

\ncmd{\ba}{\bar \alpha}
\ncmd{\bnabla}{\bar \nabla}

\ncmd{\Td}{T^{d+2}} 
\ncmd{\txi}{\tilde \xi} 
\ncmd{\bs}{\hat {\bf s}}

\ncmd{\nn}{\nonumber \\}

\newcommand{\bqa}{\begin{eqnarray}} 
\newcommand{\eqa}{\end{eqnarray}}

\definecolor{new_color}{RGB}{50,155,0}

\ncmd{\vrho}{\varrho}

\makeatletter
\newcommand*{\rom}[1]{\expandafter\@slowromancap\romannumeral #1@}
\makeatother

\begin{document}

\title{
Emergent time and more from wavefunction collapse in general relativity 
}

\author{Sung-Sik Lee\\
\vspace{0.3cm}
{\normalsize{Department of Physics $\&$ Astronomy, McMaster University,}}
{\normalsize{1280 Main St. W., Hamilton ON L8S 4M1, Canada}}
\vspace{0.01cm}\\
{\normalsize{Perimeter Institute for Theoretical Physics,
31 Caroline St. N., Waterloo ON N2L 2Y5, 
Canada}}
}

\date{\today}

\begin{abstract}
In this paper, we further develop a recently proposed theory of time based on wavefunction collapse in general relativity.
It is based on the postulations that quantum states, which violate the momentum and Hamiltonian constraints, represent instances of time, and stochastic fluctuations of the lapse and shift generate the time evolution under which an initial state gradually collapses toward a diffeomorphism-invariant state. 
Under the wavefunction collapse, the scale factor monotonically increases, thus acting as a clock.
The scalar, vector, and tensor gravitons arise as physical excitations, and the arrow of time for their evolution is set by the initial state.
In the long-time limit, the tensor gravitons exhibit emergent unitary dynamics. 
However, the extra modes are strongly damped due to the non-unitary dynamics that suppress the constraint-violating excitations.
The vector mode is uniformly suppressed over all length scales, but the decay rate of the scalar is proportional to its wave vector.
This makes the latter a viable candidate for dark matter; excitations with large wavelengths survive over long periods, contributing to long-range interactions, while the fast decay of short-wavelength modes renders them undetectable without sufficient temporal resolution.  
These are demonstrated for the cosmological constant-dominated universe through semi-classical and adiabatic approximations, which are controlled in the limit of large space dimension. 
\end{abstract}

\maketitle

\section{Introduction}

In the standard approach to quantum gravity\cite{PhysRev.160.1113}, physical states must be invariant under spacetime diffeomorphisms.
Since such a state represents a whole spacetime, constructing a history out of a single state requires a prescription to unpack a series 
of temporally separate events\cite{ 1992gr.qc....10011I, 1992grra.conf..211K, https://doi.org/10.1002/andp.201200147}.
One way of extracting what may be regarded as time evolution is through the use of correlations among dynamical variables\cite{PhysRevD.27.2885,PhysRevD.65.124013,Dittrich2007}.
One can define an instance of time through a projective measurement of a dynamical variable chosen as a clock.
The dynamics of other variables relative to the clock are encoded in the conditional probability that depends on the measurement outcome of the clock variable.
However, this relational interpretation of time has some conceptual difficulties.
First of all, there are many different ways to define an instance of time. 
The `phenomenon' of many-fingered time, which is already present in classical general relativity\cite{PhysRev.126.1864}, is elevated to another level in quantum gravity. 
Suppose a time slice is defined by a set of local clocks distributed in space, and one defines the geometry of space at the moment when all local clocks read a certain value.
Alternatively, one can consider a different set of local clocks related to the former through a non-on-site (non-ultra-local) unitary transformation. 
In general, the space that arises at an instance defined by the new clocks has a different local entanglement structure, even a different space dimension\cite{Lee:2021ta}.
Even if it is agreed to use a specific set of clocks, the correlation does not necessarily provide a chronological order.
The correlation determines the probable location of Earth as a function of the universe's size but does not explain why the universe expands.
Furthermore, the reduced quantum state obtained from a projective measurement of the clock is not invariant under diffeomorphism.
If one wants to define a notion of instances of time, it seems inevitable to consider states that break diffeomorphism invariance.

There have been various attempts to extract time in quantum gravity\cite{ 2008arXiv0808.1223B,
time_timeless,
Horwitz:1988aa,
ISHIBASHI1997467,
Connes:1994aa,
SMOLIN201586,
Carroll:2004pn,
MAGUEIJO2021136487,
Hull:2014aa,
Lee:2020aa,
PhysRevD.108.086020,
Brahma:2022aa}.
Here, we pursue a recent proposal based on wavefunction collapse\cite{glfz-yvnl}.
It is built upon two postulates.
Firstly, the physical Hilbert space is enlarged to include states that are not invariant under diffeomorphism.
Naturally, an instance of time is represented by a {\it physical} state in the extended physical Hilbert space.
Secondly, the time evolution corresponds to a process in which a diffeomorphism non-invariant initial state gradually collapses toward a diffeomorphism-invariant state.
In some approaches to quantum mechanics that aim to address the measurement problem, non-unitary components are introduced within the fundamental time evolution\cite{
PhysRevD.34.470,
PhysRevA.42.78,
PhysRevA.40.1165,
Penrose:1996aa,
2025JPhCo...9f5001F}.
In this proposal\cite{glfz-yvnl}, the non-unitary wavefunction collapse constitutes the entire time evolution.
Accordingly, states at early times have relatively large violations of the constraints associated with diffeomorphism.
The constraints are enforced only asymptotically in the long-time limit, as the time evolution gradually projects out components that are not invariant under diffeomorphism.
For an infinite-dimensional Hilbert space, the time variable, which is conjugate to the Hamiltonian, is non-compact, and the process of wavefunction collapse never ceases at any finite time.

As the first step, the proposal has been applied to the mini-superspace model of cosmology\cite{glfz-yvnl}.
In that model, the simplicity arises because the diffeomorphism is reduced to the group generated by a single operator, the Hamiltonian.
The wavefunction collapse is induced through a stochastic fluctuation of the lapse. 
While the resulting dynamics are intrinsically non-unitary and non-directional, a sense of unitarity and directionality emerges in the late time limit.
Unitarity emerges because the norm of a state decreases in a manner that is largely independent of the state in the long-time limit.
This allows one to normalize the wavefunction in a state-independent way, which effectively restores unitarity.
The directionality originates from the asymmetry of the phase space in which the stochastic fluctuations of the lapse induce a random walk.
Since the Hilbert space is asymmetric between configurations with large and small scale factors, the random walk causes the state to evolve preferentially toward one direction over the other.
The scale factor monotonically increases with projection because it plays the role of a configuration-dependent effective mass, and the state tends to be trapped in the region of slow dynamics associated with large effective masses.

While the cosmological model is a useful tool for testing the principle in a minimal setting, there are limitations.
First and foremost, general relativity has infinitely many local constraints, and it is not clear how the wavefunction collapse generated by one constraint should be extended to the infinite-dimensional case.
Furthermore, it remains to be seen whether the emergent unitarity and directionality continue to hold in general relativity, which has propagating degrees of freedom.
In particular, it is curious to know whether or not unitarity emerges uniformly across all physical degrees of freedom, and how the predicted dynamics compare with the standard predictions of general relativity as an effective field theory\cite{PhysRevD.50.3874,Burgess:2004aa}.

The goal of this paper is to extend the earlier proposal to full-fledged general relativity and to identify the dynamics of the propagating modes.
In general relativity, the symmetry includes spatial diffeomorphism as well as temporal diffeomorphism. 
Given that they form a closed algebra\cite{PhysRev.116.1322,TEITELBOIM1973542}, it is natural to treat them on an equal footing.
Therefore, the physical Hilbert space is enlarged to include states that violate both the Hamiltonian and momentum constraints.
Accordingly, the time evolution is generated by stochastic fluctuations of the lapse and shift functions, which cause the state to collapse toward a diffeomorphism-invariant state.
We confirm that the directionality emerges in the same manner as it does in the mini-superspace model:
due to the asymmetry of the Hilbert space in the direction of the scale factor, the random walk causes the scale factor to increase during the wavefunction collapse.
With the extension of the physical Hilbert space, general relativity has both a scalar mode and a vector mode in addition to the tensor gravitons as physical degrees of freedom.
Their dynamics are intrinsically non-unitary because the entire time evolution is generated by the collapse of the wavefunction.
Nonetheless, the tensor graviton exhibits emergent unitarity in the late time limit, as its decay rate relative to the real part of the energy decreases with increasing time.
This is because the tensor graviton describes excitations within the sub-Hilbert space that satisfy the constraints, and those excitations are not strongly affected by the non-unitarity.
On the other hand, the scalar and vector modes, which are the constraint-violating excitations, are strongly damped at late times.
The non-unitary dynamics of the extra modes are the key difference of this theory compared with other theories of gravity that support such extra degrees of freedom\cite{ PhysRev.124.925, Gasperini:1987aa, PhysRevD.79.084008, PhysRevD.70.083509, PhysRevD.64.024028, Moffat_2006, Pirogov:2015aa,
PhysRevLett.104.181302}. 
While the vector mode is uniformly damped at all length scales, the scalar mode is only marginally damped.
Namely, the imaginary part of the energy for the scalar mode is proportional to the real part, and its lifetime is proportional to its wavelength.
As a result, at long distances, the scalar can have prominent observational effects that differ from those predicted by the standard theory based on the strict constraints.

\section{Wavefunction collapse from stochastic fluctuations of lapse and shift}

In general relativity, the generators of the $(d+1)$-dimensional spacetime diffeomorphism are written as
\bqa
\hat  {\cal H}(x)  &=&
\BL
\frac{ \hGmnrs }{\hsg}
\hPimn
\hPirs
+ 
\hsg
\left\{
\Lambda_0
- \hat R \right\}
\BR,  
\nn
\hat  {\cal P}^\mu(x)  
&=&
\BL 
- 2 
\nabla_\nu \hPimn 
\BR.
\label{eq:HP}
\eqa
Here,
$\hgmn$ is the spatial metric, $\hg$ is its determinant
and
$\hGmnrs = 
\left(
\hat g_{\mu \rho} \hat g_{\nu \sigma}
- \frac{1}{d-1} 
\hat g_{\mu \nu} \hat g_{\rho \sigma}
\right)$.
$\hPimn$ is the canonical momentum of $\hgmn$.
$\hat R$ is the $d$-dimensional scalar curvature.
$\Lambda_0$ represents the vacuum energy.
Here, the Planck scale is set to be $1$.
All objects with `hat' are operators,
and the metric and the momentum satisfy
$[ \hPirs(x), \hgmn(y) ]
= - \frac{i}{2}(
\delta^\rho_\mu
\delta^\sigma_\nu
+
\delta^\rho_\nu
\delta^\sigma_\mu
) \delta^d(x-y)$.
For composite operators, we use 
$
\llbracket \hat {\cal O} \rrbracket 
\equiv \frac{1}{2} ( \hat {\cal O} + \hat {\cal O}^\dagger)$
to symmetrize.
Symmetrized operator
$\llbracket \hat {\cal O} \rrbracket$ satisfies
$\langle  
\llbracket \hat {\cal O}  \rrbracket
\Psi_1, \Psi_2 \rangle = \langle \Psi_1,  
\llbracket \hat {\cal O}  \rrbracket
\Psi_2 \rangle$
for $\Psi_1, \Psi_2$ 
that decay sufficiently fast at $g=0$ and $\infty$,
where $\langle 
\Psi_1, \Psi_2 \rangle
= \int Dg
\Psi_1^*(g) 
\Psi_2(g)
$ represents the inner product.
For general square integrable wavefunctions, however,
$\langle  \hat {\cal H} \Psi_1, \Psi_2 \rangle 
\neq \langle \Psi_1,  \hat {\cal H} \Psi_2 \rangle$ due to the singularity of $\hat {\cal H}$ at $g=0$ and $\infty$\cite{glfz-yvnl}.
Therefore, $\llbracket \hat {\cal H} \rrbracket$ is not essentially self-adjoint.

\begin{figure}
    \centering
\includegraphics[width=\linewidth]{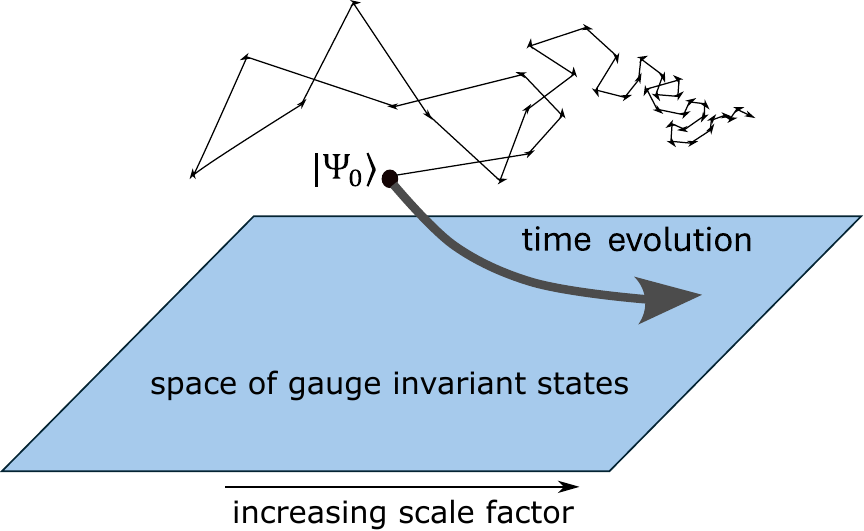} 
\caption{
In this theory, an initial state $|\Psi_0 \rangle$ does not satisfy the momentum ($\hat {\cal P}^\mu$) and Hamiltonian ($\hat {\cal H}$) constraints.
An evolution is generated by the $\hat {\cal H}$ and $\hat {\cal P}^\mu$ with a series of random lapse and shift functions.
An ensemble of such random walks induces a gradual collapse of the state toward a diffeomorphism-invariant state.
The stochastic evolution causes the state to evolve toward the direction of increasing scale factor due to the slow dynamics in the region of the Hilbert space with the large scale factor.
}
\label{fig:stochastic}
\end{figure}

Our starting point is the assumption that the states in the physical Hilbert space ${\mathbb H}_{phys}$ are not necessarily invariant under the diffeomorphism:
a state $|\Psi_0\rangle$ in ${\mathbb H}_{phys}$ generally violates the Hamiltonian and momentum constraints:
\bqa
\hat  {\cal H}(x)  
|\Psi_0 \rangle \neq 0, ~&&~
\hat  {\cal P}^\mu(x)  
|\Psi_0 \rangle \neq 0.
\eqa
The Hilbert space ${\mathbb H}_0$,
which is composed of states that obey the constraints
is a proper subspace of ${\mathbb H}_{phys}$.
In our proposal, the time evolution corresponds to a process in which an initial state 
in ${\mathbb H}_{phys}$  gradually collapses toward a state in 
${\mathbb H}_0$.
To describe such a collapse, we first consider a series of lapses and shifts that make the initial state $|\Psi_0\rangle$ evolve as
\bqa
&& | \Psi( \epsilon^{(N)}, \epsilon^{(N-1)}, .. \epsilon^{(1)} ) \rangle
= \nn &&
e^{-i  \hat H[\epsilon^{(N)}] }~
e^{-i  \hat H[\epsilon^{(N-1)}] }
\cdots
e^{-i  \hat H[\epsilon^{(1)}] }
|\Psi_0 \rangle,
\label{eq:Psiepsilons0}
\eqa
where
$\hat H[\epsilon]
= 
\int d^dx~  
\left[
\epsilon_0(x) \hat {\cal H}(x)
+
\epsilon_\mu(x) \hat {\cal P}^\mu(x)
\right]$.
It is straightforward to write
\eq{eq:Psiepsilons0} as the path-integration of the phase space variables.
For this, we first rewrite $\hat {\cal H}$ in the normal-ordered form in which $\hgmn$'s are placed to the left of $\hPirs$'s, 
\bqa
\hat {\cal H}(x) = 
%
\frac{ \hGmnrs }{\hsg}
\hPimn
\hPirs
-
i A_0
\frac{1}{\sqrt{\hg} }
\hat \Pi
+ 
\hsg
(\Lambda_0  - \hat R ).
\label{eq:He2}
\eqa
Here, $\hat \Pi =  \hat g_{\mu \nu} \hat \Pi^{\mu \nu}$
and
$A_0 = \frac{d^2-d-3}{2(d-1)}
\delta(0)$.
The divergence in $A_0$ arises from the commutator between $\hgmn$ and $\hPimn$ evaluated at the same points in space. 
In a UV-complete theory, it is expected to be regularized to a finite value. 
The precise value of $A_0$ also depends on how ${\cal H}$ is symmetrized. 
Here, we view general relativity as a long-distance effective theory and treat $A_0$ as a finite parameter of the theory.
Additional terms independent of $\Pimn$ have been absorbed into the cosmological constant.
In the path-integral, \eq{eq:Psiepsilons0} is written as
$| \Psi( \epsilon^{(N)}, \epsilon^{(N-1)}, .. \epsilon^{(1)} ) \rangle
=
\int 
{\cal D} g  {\cal D} \Pi 
~
| g^{(N)} \rangle ~
e^{i S} ~
\Psi_0(g^{(0)})$,
where
${\cal D} g \equiv \prod_{m=0}^N Dg^{(m)}$
and
${\cal D} \Pi \equiv \prod_{m=1}^N D\Pi^{(m)}$,
and 
\bqa
S &= &
\sum_{m=1}^N 
\int dx
\Bigg\{ 
\Pi^{(m)\mu \nu}
( g^{(m)}_{\mu \nu} - g^{(m-1)}_{\mu \nu} )
+ 2 \epsilon^{(m)}_{\mu} 
\nabla_\nu \Pi^{(m) \nu \mu} 
\nn &&
- \epsilon^{(m)}_0
\Bigg[
\frac{
\Gmnrs^{(m)}
}{
\sqrt{g^{(m)}}
}
\Pi^{(m) \mu \nu} 
\Pi^{(m) \rho \sigma} 
- i  A_0
\frac{\Pi^{(m)}}{ \sqrt{g^{(m)}} }
\nn && 
~~~~~~~~~~
+
\sqrt{g^{(m)}}
( \Lambda_0 -  R^{(m)} )
\Bigg]
\Bigg\}.
\label{eq:pathintegral}
\eqa
In obtaining this expression, it is convenient to use the normal ordering in \eq{eq:He2}.
In different normal orderings, $A_0$ is modified, but one also needs to include additional boundary terms due to the essentially non-self-adjoint nature of $\hat {\cal H}$
(see Appendix \ref{app:1} for details).
To describe a gradual collapse of the initial state toward ${\mathbb H}_0$, we write the infinitesimal gauge parameters as
$ \epsilon_0^{(m)}(x) = \varepsilon n_0^{(m)}(x) $
and
$ \epsilon_\mu^{(m)}(x) = \varepsilon n_\mu^{(m)}(x) $.
Here, $\varepsilon$ is an infinitesimal parameter, and $n_0^{(m)}(x)$ and $n_\mu^{(m)}(x)$ are, respectively, the lapse and shift functions. 
The collapse arises from fluctuations of the lapse and shift functions,
\bqa
|\Psi \rangle 
&=&
\int  
{\cal D} g  
{\cal D} \Pi 
| g^{(N)} \rangle 
\left[
\int
{\cal D} n  
 \rho(n) 
e^{i S} 
\right]  
\Psi_0(g^{(0)}), 
\label{eq:randomwalk}
\eqa
where 
$\int {\cal D} n  =
\int \left[  \prod_{m=1}^N    D n^{(m)}   \right]$ 
averages over the lapse and shift functions over the distribution $\rho(n)$.
To realize an unbiased collapse that does not have a preferred direction, we demand that $\rho(n)$ is even in $n_0$ and $n_\mu$.
We also assume that the lapse and shift have zero correlation length in space.
We write the second moments of the distribution as
$\int {\cal D} n \rho(n) n^{(m)}_0(x)  n^{(m')}_0(y)  = \frac{2   \xi^{(m)}(x) }{\varepsilon \sqrt{g^{(m)}}} \delta(x-y)\delta_{m,m'} $, 
$\int {\cal D} n \rho(n) n^{(m)}_\mu(x) n^{(m')}_\nu(y)  = \frac{ \xi^{'(m)}(x)g^{(m)}_{\mu\nu} }{2 \varepsilon \sqrt{g^{(m)}}} \delta(x-y) \delta_{m,m'} $ 
and 
$\int {\cal D} n \rho(n) n^{(m)}_0(x) n^{(m')}_\nu(y)  = 0$.
Here,
$\xi^{(m)}(x)$ and $\xi^{'(m)}(x)$ are functions that characterize the speed of collapse, which is generally position-dependent.
The average over the lapse and shift gives
\bqa
&& \int  {\cal D} n  
~ \rho(n)  ~
e^{- i \varepsilon \sum_m \int dx \left[  n_0^{(m)} {\cal H}^{(m)}
+ n_\mu^{(m)} {\cal P}^{(m)\mu}
\right]} = \nn
&&
\exp\Bigg\{ 
- \sum_m \int dx 
\frac{\varepsilon}{\sqrt{g^{(m)}}} 
\Bigg[ \xi^{(m)}  
({\cal H}^{(m)})^2
+ \xi^{'(m)} 
({\cal P}^{(m)})^2
 \nn &&
 ~~~~~~~~~~
 + 
 {\cal O} \left( 
 \frac{({\cal H}^{(m)})^4}{ g^{(m)} }  
 \right) +
 {\cal O} \left( 
 \frac{({\cal P}^{(m)})^4}{ g^{(m)} }  
 \right)
\Bigg]
\Bigg\},
\eqa
where 
$({\cal P}^{(m)})^2
=g^{(m)}_{ \mu\nu} {\cal P}^{(m) \mu} {\cal P}^{(m) \nu}$,
and
the last two terms are the non-Gaussian contributions. 
For an initial state that strongly violates the constraints, the higher order terms in 
${\cal H}$
and
${\cal P}$ are important initially.
However, they become less important at later times as the violation of the constraints monotonically decreases over the course of the collapse.
Here, we focus on the evolution dominated by the quadratic terms in 
${\cal H}$ and ${\cal P}$,
and ignore the non-Gaussian contributions.
In the large $N$ and small $\varepsilon$ limits with fixed $L \equiv N \varepsilon$,
we obtain a $L$-dependent state given by
%
\bqa
|\Psi(L) \rangle
=
\int 
{\cal D} g 
{\cal D} \Pi 
~~
| g_f \rangle ~
e^{i S'} ~
\Psi_0(g_i),
\label{eq:PsiL}
\eqa
where 
$g_{i, \mu \nu} = g_{\mu \nu}(0)$,
$g_{f, \mu \nu} = g_{\mu \nu}(L)$,
and
$S' =  \int_0^{L} dl \int dx \left\{ \Pi^{\mu \nu} \partial_l g_{\mu \nu} +i  \frac{1}{\sg} \left[ \xi  
{\cal H}^2
+ \xi' {\cal P}^2 \right]
\right\}$
with
${\cal P}^2 = {\cal P}^\mu {\cal P}_\mu $.
%
%
%
%
With a shift of the conjugate momentum,
$
\Pi^{\mu \nu} \rightarrow
\Pi^{\mu \nu} 
- i \frac{d-1}{2} A_0 g^{\mu \nu}
$, the action becomes
\bqa
S' &=&
\int_0^{L} dl \int dx
\Bigg\{ 
\Pi^{\mu \nu}
\partial_{l}
g_{\mu \nu}
- i \frac{d-1}{2} A_0
\partial_{l} \log g 
\nn &&
+i  \frac{\xi}{\sg}
\left[
\frac{ \Gmnrs }{\sg}
\Pi^{\mu \nu} 
\Pi^{\rho \sigma} 
+ \sg ( \Lambda -  R  )
\right]^2 \nn &&
+ i \xi' \frac{g_{\mu \nu}}{\sg} 
\nabla_\rho \Pi^{\mu \rho}
\nabla_\sigma \Pi^{\nu \sigma}
\Bigg\},
\label{eq:fullS}
\eqa
where 
$\Lambda = \Lambda_0 - \frac{d(d-1)}{4 g} A_0^2$.
We may well take \eq{eq:fullS} as the starting point of our theory.

For constant $\xi$ and $\xi'$,
$L$ is related to the violation of the constraints through
$L \sim 
\left[ 
\int dx ~
\langle \Psi(L) |
\frac{1}{\hsg}
\left[ 
\xi \hat {\cal H}^2
+  \frac{\xi'}{4} \hat {\cal P}^2
\right]
| \Psi(L) \rangle
\right]^{-1}$.
With increasing $L$, the initial state  collapses toward the direction of decreasing
$\hat {\cal H}^2$ and  $\hat {\cal P}^2$.
We interpret $L$ as time, and $|\Psi(L) \rangle$ describes the evolution of the quantum state as a function of time.
Although the phase space path integration formally takes the form of a Euclidean theory,
\eq{eq:fullS} does not have the conformal problem of Euclidean gravity\cite{GIBBONS1978141} because ${\cal H}$ and ${\cal P}^\mu$ have been squared and the imaginary part of the action is strictly non-negative. 
Besides $\hat {\cal H}^2$ and  $\hat {\cal P}^2$, the action also has a total derivative term, 
$- i \frac{d-1}{2} A_0 \partial_{l} \log g$.
It gives boundary terms at $l=0$ and $L$, contributing 
$e^{(d-1)A_0 \int dx \log g}$
to the probability distribution 
$|\langle g |\Psi(L) \rangle|^2$ as a function of $g$.
This boundary term shifts the probability toward configurations with large $g$. 
This term plays an important role in determining the average size of the universe as a function of $L$, as will be shown later.
It also has a preferred spatial coordinate, the coordinate in which $A_0$ is constant.
While the spatial diffeomorphism is generally broken by the initial state and the boundary action, the bulk action itself respects it.
Since the terms in the bulk action satisfy the first-class algebra,  $|\Psi(L)\rangle$ approaches a well-defined state in the large $L$ limit.

\section{Arrow of time}
\label{sec:arrow}

\begin{figure}
    \centering
\includegraphics[width=0.9\linewidth]{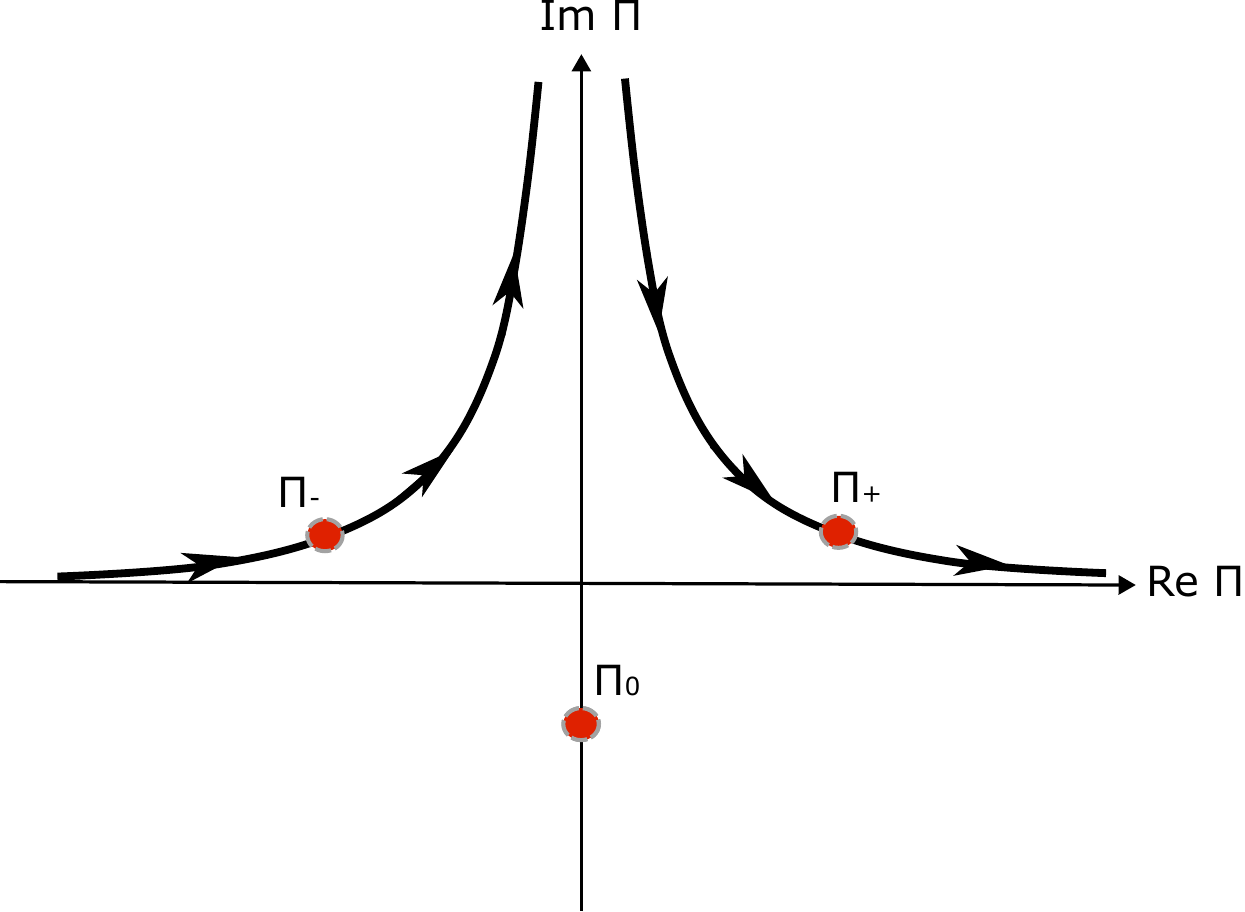} 
\caption{
The original contour of $\Pi$ runs along the real axis. 
It can be deformed to the path of the stationary phase (denoted as the solid line) that goes through two saddle-points, $\Pi_\pm$.
}
\label{fig:contour}
\end{figure}

Because the diffeomorphism is not a gauge symmetry, the physical degrees of freedom include a scalar mode and a vector mode, in addition to the tensor gravitational mode. 
Having identified $L$ as time, our goal is to compute $|\Psi(L)\rangle$ as a function of $L$ and to extract the effective theory that governs the dynamics of the physical degrees of freedom.
Henceforth, we focus on the case where the evolution is dominated by a positive cosmological constant ($\Lambda>0$).
We first single out the scalar mode associated with the scale factor and its conjugate momentum by writing $g_{\mu \nu}$ and $\Pi^{\mu \nu}$ as 
\bqa
\gmn =  e^{2 \alpha} \tilde g_{\mu \nu},  ~~~~
\Pimn  = 
\left( \frac{\Pi}{d} g^{\mu \nu} +\tilde \Pi^{\mu \nu}\right).
\eqa
Here, $\alpha$ is the logarithmic scale factor and $\tilde g_{\mu \nu}$ is unimodular metric. 
$\Pi= \Pimn \gmn$. 
$\tilde \Pi^{\mu \nu}$ is the traceless part of the conjugate momentum.
In terms of these variables, the action is written as
\begin{widetext}
\bqa
S_0 &= &
\int_0^{L} dl \int dx
\left\{ 
2 \Pi \partial_l \alpha 
+ e^{2\alpha} \tilde \Pi^{\mu \nu} \partial_l \td g_{\mu \nu}
- i A  
\partial_l \alpha
+i 
e^{d\alpha}
\txi
\left[ 
\Pic^2
-
e^{-2d\alpha}
\Pi^2
\right]^2  
\right. \nn && \left.
+ i \xi'
e^{-d \alpha}  
\left[
\frac{1}{d^2} \nabla_\mu \Pi \nabla^\mu \Pi
+
\frac{2}{d} \nabla_\mu \Pi \nabla_\nu \tilde \Pi^{\nu \mu}
+ g_{\mu \nu} \nabla_\rho \td \Pi^{\mu \rho} \nabla_\sigma \td \Pi^{\nu \sigma}
\right]
\right\},
\label{eq:Salpha0}
\eqa
\end{widetext}
where
$\txi = \frac{\xi}{u^2}$
and
$A  = u A_0$
with
$u=d(d-1)$
and
\bqa
\Pic^2 = 
u
\Bigl[
\Lambda
+
e^{-2d\alpha}
\Gmnrs
\td \Pi^{\mu \nu} 
\td \Pi^{\rho \sigma} 
-  R 
\Bigr].
\label{eq:h}
\eqa
From now on, we choose $\tilde \xi=\xi'=1$. 
Being a quartic function of $\Pi$,
$S_0$  supports three saddle points for $\Pi$.
For fields that are slowly varying in $l$ and $x$,
the saddle-points for $\Pi$ can be written as 
\bqa
\Pi_0 =  
-  i 
e^{d\alpha} 
\frac{\partial_l \alpha}{2  \Pic^2},  ~~~
\Pi_\pm = 
e^{d\alpha} 
\left(
\pm \Pic
+ i \frac{\partial_l \alpha}{4  \Pic^2}
\right)
\label{eq:saddlepts}
\eqa
to the leading order in $\partial_l$ and $\nabla$, where
$\Pic = 
\sqrt{
u \Lambda
}
\left[ 
1
+ 
\frac{1}{2}
\frac{ \Ht }{\Lambda}
+ 
O\left(
\left(
 \Ht /\Lambda
\right)^2
\right)
\right]$
with
\bqa 
\Ht=
e^{-2d\alpha} \Gmnrs
\td \Pi^{\mu \nu} \td \Pi^{\rho \sigma}  -  R.
\label{eq:Ht}
\eqa
The original contour of $\Pi$ can be deformed to the path of a stationary phase that goes through $\Pi_\pm$
as is shown in \fig{fig:contour}.
Around saddle point $\Pi_s$ with $s=\pm$, the action can be written as 
$S_0 = S_{1}+S_{1}'$, where
\bqa
&&S_{1} = 
\int_0^{L} dl \int dx 
\Bigg\{
e^{2\alpha} \tilde \Pi^{\mu \nu} \partial_l \td g_{\mu \nu}
+ 2 s
e^{d\alpha} 
\Pic   \partial_l \alpha  
\nn && 
+i  
\left[
\frac{
e^{d\alpha} 
}{4  \Pic^2}
( \partial_l \alpha)^2
-  A   \partial_l \alpha   
\right] 
+ i  e^{-d \alpha} 
g_{\mu \nu}
\nabla_\rho 
\td \Pi^{\mu \rho}
\nabla_\sigma 
\td \Pi^{\nu \sigma}
\Bigg\},  \nn
\label{eq:Salpha}
\eqa
is the saddle point action and
\bqa
&& S_{1}' = 
i \int_0^{L} dl \int dx 
e^{d\alpha} 
\Bigg[
\sqrt{  16 P^4
+ \left( \frac{3\partial_l \alpha }{P} \right)^2 } 
\left( e^{-d\alpha} \delta \Pi \right)^2 \nn &&
~~~~~~~~~~ + O\left( ( e^{-d\alpha} \delta \Pi)^3 \right)
\Bigg]
\label{eq:Salphaprime}
\eqa
is the action for $\delta \Pi$ that represents deviation of $\Pi$ away from the saddle point along the stationary path. 
The coefficients for the cubic and quartic terms are of the order of $P$ and $1$, respectively.
In the large $d$ limit for a fixed $\Lambda > 0$, $P^2 \approx d(d-1) \Lambda$ becomes large, and the saddle-point approximation becomes exact.
Within the saddle-point approximation, we obtain
\bqa
|\Psi(L) \rangle
=
\sum_{s=\pm} \int 
{\cal D} \alpha 
{\cal D} \td g 
{\cal D} \td \Pi
~
| g_f  \rangle ~
e^{i 
S_{1}
} ~
\Psi_0(g_i),
\label{eq:PsiL2}
\eqa
where $s=\pm$ denote the contributions from $\Pi_{\pm}$.

The real part of $S_1$, the first two terms in \eq{eq:Salpha}, describes the unitary dynamics for 
$\tilde g_{\mu \nu}$ and $\tilde \Pi^{\mu \nu}$.
By rewriting them as 
\bqa
\Re~ S_1 =
\int d\alpha \int dx
\Bigg\{
e^{2\alpha} \tilde \Pi^{\mu \nu} \partial_\alpha \td g_{\mu \nu}
+ 2 s 
e^{d\alpha} 
\Pic
\Bigg\},
\eqa
we readily note that  $\alpha$ plays the role of time for the evolution of the unimodular metric,
$2 P$ is the Hamiltonian density, 
and $-s$ sets the direction of the time evolution. 
With $s$ summed over $\pm 1$, both branches of forward and backward evolution are included.
However, one branch is chosen by the initial state if it has a well-defined momentum for $\alpha$.
For $\Psi_0 \sim e^{2 i\int dx 
\pi_\alpha \alpha}$, where $\pi_\alpha$ is the conjugate momentum of $\alpha$,
the equation of motion of $\alpha(0)$ in \eq{eq:PsiL2} enforces $s e^{d\alpha} P  = \pi_\alpha$.
Since $P$ is positive, the sign of $s$ is fixed to be that of $\pi_\alpha$.
Therefore, the direction of time evolution for $\td g_{\mu \nu}$ relative to $\alpha$ is determined by the sign of the momentum of the scale factor.
The origin of the state-dependent arrow of time can be understood in the following way.
In \eq{eq:Psiepsilons0}, the state undergoes an evolution generated by 
$ e^{-i \int dx \epsilon_0(x)
e^{d\alpha} \left( 
\Pic^2 -
e^{-2d\alpha} \Pi^2 
\right)
}$
with the fluctuating lapse $\epsilon_0(x)$.
For the initial state with ${\cal H} \approx 0$ (but not exactly zero) and initial momentum $\pi_\alpha$,
$e^{-2d\alpha} \Pi^2  \approx \Pic^2$
and $\Pi \approx \pi_\alpha \approx s e^{d\alpha} P$.
For such a state, the evolution operator can be approximated as
\bqa
e^{-i \int dx \epsilon_0(x)
e^{d\alpha} s \sqrt{u\Lambda} 
\left( 
2 s \Pic - 
e^{-d\alpha} 2 \Pi 
\right) 
},
\label{eq:randomH}
\eqa
where 
$P\approx \sqrt{u\Lambda}$ is used. 
Relative to $2 \Pi$, which generates the translation of $\alpha$, the evolution of the unimodular metric is generated by $-2sP$.
This shows that $2P$ and $-s$ act as the Hamiltonian and the direction of time, respectively, for the unimodular metric relative to $\alpha$.
Henceforth, we assume that the initial state $|\Psi_0\rangle$ has $\pi_\alpha$ of sign $s$. 
Accordingly, that particular $s$ is chosen in the sum of branches in \eq{eq:PsiL2}.

In \eq{eq:randomH}, we establish that the initial momentum of $\alpha$ sets the direction of the evolution of $\td g_{\mu \nu}$ relative to $\alpha$.
However, it is not enough to understand the relative evolution between the unimodular metric and the scale factor.
In our theory, there is an absolute time $L$, and $\alpha$ is an independent dynamical variable as well.
The imaginary kinetic term for $\alpha$,
$i \frac{ e^{d\alpha}  }{4  \Pic^2} ( \partial_l \alpha)^2$ in \eq{eq:Salpha},
is the result of the `soft' Hamiltonian constraint enforced by the imaginary action $\frac{i}{\sg}{\cal H}^2$ in \eq{eq:fullS}. 
Since the Hamiltonian constraint can be violated with a complex energy penalty, $\alpha$ becomes a dynamical variable with complex kinetic energy.
If the Hamiltonian constraint was imposed strictly through 
$\delta \left( {\cal H} \right)$, 
the kinetic term
would be absent, 
and $\alpha$ would be non-dynamical. 
Similarly, the soft momentum constraint imposed by 
$\frac{i}{\sg} 
{\cal P}^2$
makes the vector mode conjugate to $\nabla_\nu \Pimn$ dynamical variables.

To understand the dynamics of all physical degrees of freedom ($\alpha$ and $\tilde g_{\mu \nu}$) relative to time $L$, we first identify the semi-classical path and then consider fluctuations around it.
For this, we first note that \eq{eq:Salpha} includes two total derivative terms, 
\bqa
S_{\partial,1} &=& \int dl dx ~\partial_l \left( \frac{2s \sqrt{u \Lambda}}{d} e^{d \alpha}  \right), \nn
S_{\partial,2} &=& -i A \int dl dx~ \partial_l \alpha.
\label{eq:Spartial}
\eqa
They give rise to boundary terms localized at $l=0$ and $L$. 
$S_{\partial,1}$ can be readily removed through a unitary transformation.
However, $S_{\partial,2}$ can not be removed through a unitary transformation and is kept as it is.
In the rotated basis,
which is defined by
$|\tilde \Psi \rangle 
= \hat U_{s} | \Psi \rangle$,
where $\hat U_{s} = e^{-i \int dx ~ \frac{2s \sqrt{u \Lambda}}{d} e^{d \hat \alpha}}$,
Eqs. (\ref{eq:PsiL2}) and (\ref{eq:Salpha}) become 
$|\tilde \Psi_s(L) \rangle
=
\int 
{\cal D} \alpha 
{\cal D} \tilde g 
{\cal D} \tilde \Pi
~
| g_f  \rangle ~
e^{i 
\tilde S_1
} ~
\tilde \Psi_{0,s}(g_i)$,
where the subscript $s$ is a reminder of the branch chosen by the initial state,
and the action is given by
\bqa
\tilde S_1 &= &
\int_0^{L} dl \int dx 
\Bigg\{ 
e^{2\alpha} \tilde \Pi^{\mu \nu} \partial_l \td g_{\mu \nu}
+  s \sqrt{\frac{u}{\Lambda}}  e^{d\alpha}  
\Ht \partial_l \alpha  
\nn && 
+i  
\left[
\frac{ e^{d\alpha}  }{4  u \Lambda} 
\left( 1 -  \frac{ \Ht }{\Lambda} \right)
( \partial_l \alpha)^2
- 
A  
 \partial_l \alpha   
\right] 
\nn && 
+ i 
e^{-d \alpha}  
g_{\mu \nu}
\nabla_\rho \td \Pi^{\mu \rho}
\nabla_\sigma \td \Pi^{\nu \sigma}
\Bigg\}.
\label{eq:tildeSalpha}
\eqa
Here, the higher-order terms in $\Ht/\Lambda$ are dropped.


\begin{figure}
    \centering
\includegraphics[width=1.0\linewidth]{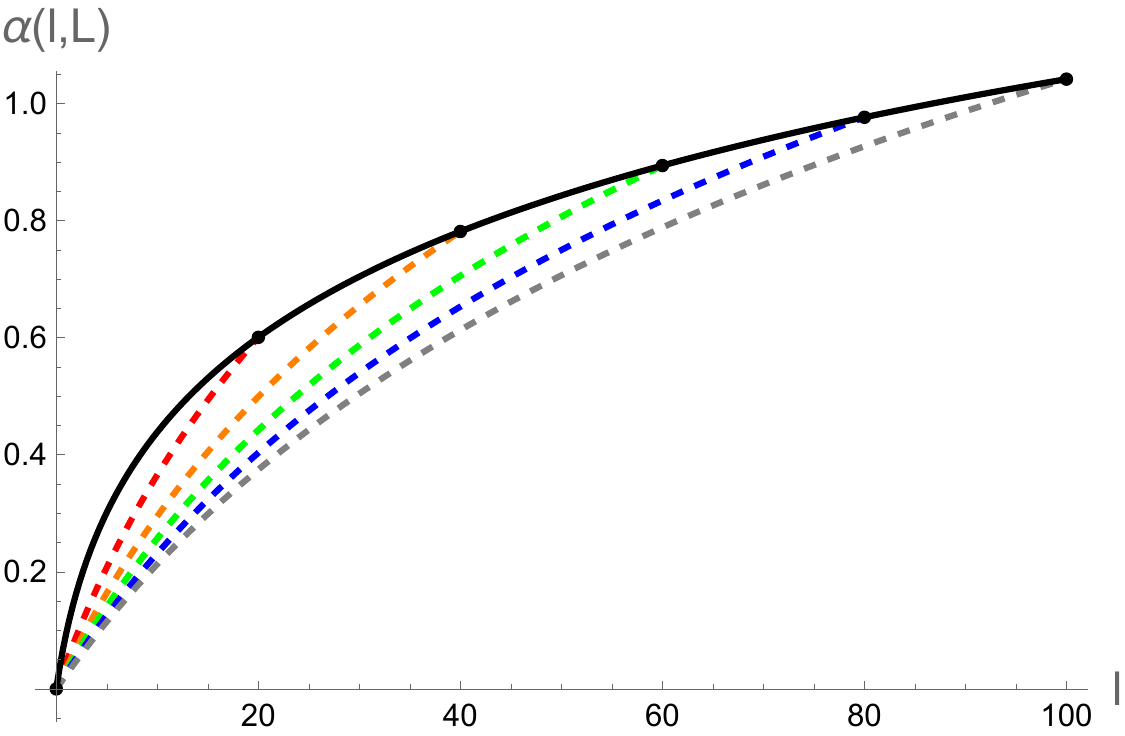} 
\caption{
The dashed lines from left to right represent
$\ba(l;L)$ plotted as a function of $l$ for $L=20,
40,
60,
80,
100$, respectively.
The solid line denotes the evolution of the scale factor as a function of $L$, $\ba(L;L)$.
For the plot, we choose $d=3$,
$A=1$,
$\Lambda=0.01$
and
$\alpha_0=0$.
}
\label{fig:Gamma_smallL}
\end{figure}

Suppose that the initial wavefunction $\tilde \Psi_{0,s}$ is peaked at $\alpha=\alpha_0$ and describes a spatially homogeneous state.
In the semi-classical limit,
the peak of the wavefunction at $l=L$ is determined by the saddle-point path that interpolates $l=0$ and $L$,
\bqa 
g_{\mu \nu}(l;L)  = e^{2 \ba(l;L)} \delta_{\mu \nu},  
~~~ 
{\td \Pi}^{\mu \nu}=0,
\label{eq:saddleAnsatz}
\eqa
where 
$\ba(l;L)$
satisfies
\bqa
2 \partial_l^2 \ba(l;L)
+
 d  
 \left(
 \partial_l \ba(l;L)
 \right)^2
= 0
\eqa
with $\ba(0,L)=\alpha_0$.
Since the saddle-point equation is a second-order differential equation, the initial condition at $l=0$ leaves one remaining constant of integration unfixed. 
We use $\alpha(L;L)$ as that constant to write the saddle-point solution as
$\ba(l;L)=
\alpha_0
+
\frac{2}{d}
\log \left[
1+
\left(e^{ \frac{d}{2} (\alpha(L,L)-\alpha_0)}-1\right)
\frac{l}{L}
\right]$.
$\ba(L,L)$ is ultimately fixed by minimizing the imaginary part of the saddle-point action,
\bqa
\Im \Gamma=
-A  \left[ 
\ba(L,L)-\alpha_0 
\right]
+
\frac{1}{ d^2 u \Lambda   L  }
\left[
e^{\frac{d}{2} \ba(L,L)} -
e^{\frac{d}{2} \alpha_0}
\right]^2.
\label{eq:ImGamma}
\eqa
$\Im \Gamma$ is minimized for
$e^{\alpha(L,L) } =   
\left[ \frac{
e^{\frac{d}{2} \alpha_0 }    + \sqrt{ e^{d \alpha_0 }    + 4 u d \Lambda   A   L }}{2}
\right]^{\frac{2}{d}}$,
and the saddle-point path that connects $l=0$ and $L$ is 
\bqa
\ba(l;L)=
\alpha_0
+
\frac{2}{d}
\log \left[
1+
\frac{\left( \sqrt{ 1+ 4A u d \Lambda L e^{ -d \alpha_0} } -1 \right)l}{2L}
\right].
\label{eq:balL}
\eqa
At large $l$,
$\partial_l \ba(l;L)  \sim \frac{1}{l}$, and the gradient expansion used in \eq{eq:saddlepts} is valid.
It is noted that $\ba(l;L)$ depends on both $l$ and $L$ because $\ba$ takes different saddle-point paths for different $L$, as shown in \fig{fig:Gamma_smallL}.
This is due to the fact that the saddle-point path is determined by minimizing \eq{eq:ImGamma} which includes the boundary action
$S_{\partial,2}$.
What drives the expansion of space is that boundary action:
if $A=0$, there is no expansion.
We emphasize that $\ba(l;L)$ as a function of $l$ does not represent the time evolution.
The actual time evolution of the scale factor is given by $\ba(L,L)$ as a function of L.
At large $L$,
the scale factor 
grows as 
\bqa
e^{\ba(L;L)} =
\left[ Aud \Lambda  L \right]^{\frac{1}{d}} + O(L^{1/d-1/2}).
 \label{eq:baLL}
 \eqa
It is noted that the classical scale factor increases with $L$ irrespective of $s$. 
This implies that, on average, the sign of the lapse $\epsilon_0$ in \eq{eq:randomH} is chosen to be $-s$ such that the scale factor increases under the stochastic evolution\footnote{
The scale factor increases when the lapse is chosen with the opposite sign of its momentum because the kinetic term of $\alpha$ has the wrong sign: ${\cal H}$ is proportional to $-\Pi^2$, not $\Pi^2$.}.

\section{Excitations}

With increasing $L$,
the state undergoes the collapse
and $\alpha(L;L)$ at which
$\langle g |\tilde \Psi_s(L) \rangle$ is peaked monotonically increases.
We now examine fluctuations around the saddle-point.
Small fluctuations of the metric around the saddle point are parameterized by 
a scalar $\da$, a vector $\gamma_\mu$ and a transverse and traceless tensor $h_{\mu \nu}$ as
\bqa
g_{\mu \nu} &=& 
\bar g_{\mu \nu} + 
2  \da  \bgmn
+ \bnabla_{[\mu} \gamma_{\nu]}
- \frac{ \bnabla \cdot \gamma}{d} \bgmn
+ h_{ \mu \nu},  
\eqa
where
$\bnabla_\mu = \partial_\mu$ is the covariant derivative defined with respect to the saddle-point metric $\bgmn$
in \eq{eq:saddleAnsatz}
and
$ \bgimn  h_{\mu \nu} =  \bnabla^\mu  h_{\mu \nu} =0 $.
Indices of the fluctuating fields are raised or lowered with $\bgimn$ and $\bgmn$.
With this parameterization, the determinant of $\gmn$ is written as $g=e^{2d (\ba + \da + \varphi_2)}$,
where
\bqa
\varphi_2
=
-\da^2 + \frac{1}{4d} \gamma^\mu \cM_{\mu \nu} \gamma^\nu
-\frac{1}{4d} 
h_{\mu\nu}
h_{}^{\mu\nu}
\label{eq:varphi2}
\eqa
 to the quadratic order in the fluctuating field,
and
\bqa
\cM_{\mu\nu} 
=
\frac{
\bgmn
}{2} \bnabla^2 + \frac{d-2}{2d} \bnabla_\mu \bnabla_\nu. 
\eqa
Total derivative terms in $\bnabla$ are dropped as they do not affect the Gaussian action.
Similarly, the traceless part of the conjugate momentum is written as
\bqa
\tilde \Pi^{\mu \nu}  &=&  \frac{\delta \Pi}{d} \bgimn + \bnabla^{[\mu} p^{\nu]} - \frac{ \bnabla \cdot p}{d} \bgimn + \Pi_\perp^{\mu \nu}
\eqa
with
$ \bgmn \Pi_\perp^{\mu \nu}  =\bnabla_\mu \Pi_\perp^{\mu \nu} =0 $. 
We have included $\delta \Pi \bgimn$ because 
$\td \Pi^{\mu \nu}$ is traceless with respect to the full metric $\gmn$ but not necessarily with respect to $\bgmn$.
To enforce $\td \Pi^{\mu \nu} g_{\mu \nu} =0$,
$\delta \Pi$ needs to be fixed to 
$\delta \Pi = 
p^\mu 
\cM_\mu^{\nu} 
\gamma_\nu - \Pi_\perp^{\mu \nu} h_{ \mu \nu}$
up to a total derivative term
and terms that are cubic or higher 
in the fluctuation fields.
To the quadratic order, the state at time $L$ can be 
as 
\bqa
 | \tilde \Psi_s(L) \rangle 
&=&
\int 
{\cal D} \da
{\cal D} \gamma
{\cal D} h
{\cal D} p
{\cal D} \Pi_\perp
~
| g_f \rangle ~
e^{i S_\partial(L)} 
\times \nn && 
e^{ 
i \int_0^{L} dl \int dx 
\xmd
(\bar {\cal L} + {\cal L}_2')
} ~
e^{-i S_\partial(0)}
\tilde \Psi_{0,s}(g_i).
\label{eq:PsiL3}
\eqa
Here, ${\cal \bar L} =  i   \left[ \frac{ 1 }{4 u \Lambda  } ( \partial_l {\bar \alpha})^2 -  A   \xd  \partial_l {\bar \alpha }   \right]$
is the on-shell Lagrangian density.
The quadratic Lagrangian density for the fluctuating fields reads
\begin{widetext}
\bqa
{\cal L}'_2 & = &
\xd \left\{ \Pi_\perp^{\mu \nu} 
\left[\partial_l 
h_{ \mu \nu}
-2 (\partial_l \ba )
h_{ \mu \nu}
\right] 
-  p^\mu 
\cM_\mu^\nu
\left[ \partial_l \gamma_\nu 
-  2 (\partial_l \ba ) \gamma_\nu \right] 
\right\}
+ i \xtd 
\bar g^{\mu \nu} \cM_\mu^\rho p_\rho
\cM_\nu^\sigma p_\sigma
+ \frac{i}{4\Lambda  u } 
\left( \partial_l \da 
\right)^2
\nn
&& 
- \EL 
\Bigg\{
\xtd
\Pi_\perp^{\mu \nu}
\Pi_{\perp \mu \nu}
- 
\xtd
p^\mu 
\cM_{\mu \nu}
p^\nu  
- \frac{1}{4} 
h_{}^{\mu \nu} \bnabla^2
h_{ \mu \nu} 
- (d-1)(d-2)  \left| \bnabla \left( \da - \frac{1}{2d} \bnabla \cdot \gamma \right) \right|^2 
\Bigg\}, 
\eqa
\end{widetext}
where
\bqa
\EL =  -s \sqrt{\frac{u}{\Lambda }}
(\partial_l \ba)
\left( 1 - i \frac{s (\partial_l \ba)}{4
(u \Lambda )^{3/2}} \right).
\label{eq:EL}
\eqa
%
$S_\partial$ is the boundary action generated from a total derivative term in $l$,
\bqa
S_\partial(l) &=&
-i \int dx 
\Big\{
 A (\da(l)+\varphi_2(l)) 
 \nn && 
-  
\frac{\partial_l  e^{d \ba(l;L)}  }{4 d \Lambda u}  
\left(d \da(l)^2+2 \da(l)+ 2 \varphi_2(l)\right) 
\Big\},
\eqa
where $\varphi_2$ is defined in \eq{eq:varphi2}.
From the saddle-point equation that minimizes \eq{eq:ImGamma},
the boundary term at $L$ can be simplified into 
$S_\partial(L)= 
i \frac{d}{2} A  \int dx 
\varphi(L)^2$.
Integrating out 
$\Pi_\perp^{\mu \nu}$
and
$p^\mu$, we obtain
\bqa
|\tilde \Psi_s(L) \rangle
&=&
\int 
{\cal D} \da
{\cal D} \gamma
{\cal D} h
~
\hat W
| g_f \rangle ~
\times \nn &&
e^{ 
i\int_0^{L} dl \int dx 
\xmd
(\bar {\cal L} + {\cal L}_3')
}  
\tilde \Psi'_{0,s}(g_i),
\label{eq:PsiL3V}
\eqa
where 
$\hat W = 
e^{-\frac{dA}{2} \int dx 
\hat \varphi^2
}$,
$\tilde \Psi'_{0,s}(g_i)
= e^{-i S_\partial(0)} \tilde \Psi_{0,s}(g_i)$,
and
\bqa
{\cal L}'_3  &= &
\frac{1}{4\EL} 
\left(\partial_l 
h_{\mu}^{~\nu}
\right)
\left(\partial_l 
h_{\nu}^{~\mu}
\right) 
+\frac{\EL}{4} 
 h_{ \nu}^{\mu}
 \bnabla^2
 h_{ \mu}^{\nu}
+ \frac{i}{4 \Lambda  u} 
\left( \partial_l \da  \right)^2 
 \nn &&
+ \frac{i}{4}
\left( \partial_l \gamma^\mu  
\right) 
S^{-1}_{\mu \nu}
\left( \partial_l \gamma^\nu  
\right) \nn &&
+ \EL  (d-1)(d-2) \left| \bnabla \left( \da - \frac{1}{2d} \bnabla \cdot \gamma \right) \right|^2
\label{eq:L3}
\eqa
with
$S^{-1}_{\mu \nu} =
\frac{1}{ \bnabla^2  -2 i \EL } 
\left[
\bgmn
\nabla^2 
-i (d-2) \EL 
 \frac{
\bnabla_\mu \bnabla_\nu}{
(d-1) \bnabla^2 
-id \EL
}
\right]$.
%
In the large $l$ limit, $\partial_l \ba$ and $\EL$ become vanishingly small so that
$S^{-1}_{\mu \nu} \approx \bgmn$\footnote{
In particular,
$\frac{\EL}{\bnabla^2} \sim 
\frac{\partial_l \ba}{e^{-2 \ba} \partial_\mu^2} 
\sim 
\left( \frac{l}{L} \right)^{2/d}
\frac{l^{(2-d)/d}}{\partial_\mu^2} 
\ll 1
$
in the large $l$ limit
for any fixed $\partial_\mu^2$
in $d > 2$.
}, and the action is simplified to
\bqa
{\cal L}'_3 & = &
\frac{1}{4\EL} 
\left[
\partial_l  h_{\mu}^{~\nu}
\partial_l  h_{\nu}^{~\mu}
+ \EL^2 
 h_{\mu}^{~\nu}
 \bnabla^2
 h_{\nu}^{~\mu}
 \right]
 \nn && 
 +
 \frac{i}{4 \Lambda  u} 
 (\partial_l \da)^2 
 + \frac{i}{4}
\bgmn
\partial_l \gamma^\mu   
\partial_l \gamma^\nu  
\nn &&
+ \EL  (d-1)(d-2) \left| \bnabla \left( \da - \frac{1}{2d} \bnabla \cdot \gamma \right) \right|^2. 
\label{eq:L4}
\eqa
There is a mixing between $\phi$ and $\bnabla \cdot \gamma$ in the kinetic term.
To diagonalize it, we decompose $\gamma^\mu$ into the longitudinal and the transverse modes as
$\gamma^\mu =   \nabla^\mu \theta  +   \gamma_\perp^\mu$, where  $\bnabla_\mu  \gamma_\perp^\mu = 0$.
In the rotated basis defined by
\bqa
\chi &= &
\da - \frac{1}{2d} 
\bnabla^2  \theta,  \nn
\Phi &= &
\frac{1}{\sqrt{u\Lambda \bnabla^2(\bnabla^2 - 4d^2u\Lambda)}} 
\left[
\frac{\bnabla^2}{2} \da
-
d u\Lambda \bnabla^2 \theta 
\right],
\eqa
the quadratic Lagrangian density is written as
\bqa
&& {\cal L}'_3  = 
\frac{1}{4\EL} 
\left[
\partial_l  h_{\mu}^{~\nu}
\partial_l  h_{\nu}^{~\mu}
+ \EL^2 
 h_{\mu}^{~\nu}
 \bnabla^2
 h_{\nu}^{~\mu}
 \right]
 \nn && 
 + i 
 (\partial_l \chi)
 \frac{d^2}{-\bnabla^2+4d^2u\Lambda} 
 (\partial_l \chi)  
+ \EL  (d-1)(d-2) \left( \bnabla \chi \right)^2 
 \nn &&
+ i ( \partial_l \Phi)^2 
 + \frac{i}{4} \bgmn \partial_l \gamma_\perp^\mu    \partial_l \gamma_\perp^\nu
\label{eq:L42}
\eqa
to the leading order in 
$\frac{ (\partial_l \bgmn)/\bgmn }{ (\partial_l \chi)/\chi } $
and
$\frac{ (\partial_l \bgmn)/\bgmn }{ (\partial_l \Phi)/\Phi } $.
We refer to $h_{\mu}^\nu$ and $\chi$ as the tensor graviton and the scalar, respectively. 
$\Phi$ and $\gamma_\perp^\mu$ together will be referred to as the vector modes, or
the longitudinal and transverse components of the vector, respectively.

The next step is to evaluate 
$|\td \Psi_s(L) \rangle$ in \eq{eq:PsiL3V} and extract the effective Hamiltonian that governs the evolution of $|\td \Psi_s(L) \rangle$ with increasing $L$.
In discussing the time evolution, it is convenient to rescale $L$ to define a new time coordinate in which the speed of the tensor graviton becomes $1$.
From \eq{eq:tildeSalpha}, we note that the real part of the action for the tensor graviton can be written as 
$s \sqrt{\frac{u}{\Lambda}} 
\int d\ba e^{d\ba} \Ht $.
This suggests that the proper time corresponds to 
$
\sqrt{\frac{u}{\Lambda}} 
\ba
$. 
So, we define our new time coordinate $t$ to be
\bqa
t = \sqrt{\frac{
u
}{\Lambda }}  \ba(L,L).
\label{eq:tL}
\eqa
From \eq{eq:baLL}
and 
$e^{\ba(L,L)} =
 e^{
 \sqrt{\frac{\Lambda}{u
 }}
 t}
$,
one can readily obtain the relation between $L$ and $t$ as
$L(t) = \frac{
e^{\sqrt{\frac{d \Lambda }{d-1}}t}
-
e^{\frac{1}{2} \left( 
d \alpha_0 + \sqrt{\frac{d \Lambda }{d-1}} t \right) } 
}{ A d^2(d-1) \Lambda }$.
We will show that the tensor graviton indeed has speed $1$ in this time coordinate.
It is noted that the background metric $\bgmn$ changes slowly as a function of $t$ in the large $d$ ($u$) limit. 
This allows us to use the adiabatic approximation for modes whose frequencies are much larger than the rate at which $\bgmn$ changes.

The effective Hamiltonian that governs the evolution of the state from $t$ to $t+dt$ is defined through
\bqa
| \td \Psi_{s}(t+dt) \rangle
= e^{i s \hat H_{eff}(t) dt}
| \td \Psi_{s}(t) \rangle,
\label{eq:Heff0}
\eqa
The factor of $s$ is singled out because the direction of the time evolution for the fluctuating degrees of freedom relative to $\ba$ is $-s$, 
as discussed in Sec. \ref{sec:arrow}.

In the adiabatic limit,
the effective Hamiltonian is obtained to be
(see Appendix \ref{app:Heff} for details)
\bqa
&& \hat H_{eff}(t) 
=
\sum_k
\Bigg[
\sum_j 
\Omega_{k}(t)
\left(
\hat a^{(j) \dagger}_{k}(t)
\hat a^{(j)}_{k}(t)
+ \frac{1}{2}
\right) 
\nn &&
~~~~~~~~~~ 
~~~~~~~~~~ 
~~~~~~~~~~ 
+
\Omega^\chi_k(t)
\left(
\hat b'_k{}^{\dagger}(t)
\hat b'_k(t)
+ \frac{1}{2}
\right)
\Bigg] \nn &&
+is \int dx \frac{1}{\sqrt{\bar g(t)}}
\left[
C_\Phi(t)
 (\hat \pi'_{\Phi})^2 
+ 
C_{\gamma_\perp}(t)
\bgimn(t)
 \hat \pi_{\gamma_\perp \mu}
 \hat \pi_{\gamma_\perp \nu}
 \right]. \nn
\label{eq:Heff}
\eqa
Here, 
$\hat a^{(j)}_{k}(t)$
$\left( \hat a^{(j)\dagger}_{k}(t) \right)$
denotes the annihilation (creation) operator of the tensor graviton of momentum $k$ and polarization $ j = 1,..,(d-2)(d+1)/2$ 
at time $t$. 
$ \hat b'_{k}(t)$ 
$\left( \hat b_{k}^{'\dagger}(t) \right)$
is the annihilation (creation) operator of the scalar particle.
$\hat \pi'_{\Phi}$ and $\hat \pi_{\gamma_\perp \nu}$ are the conjugate momentum of the longitudinal and transverse components of the vector, respectively.
In the large $t$ limit, the complex mode energies become
\bqa
\Omega_{k}(t) &=&
|k|
+ i s \Gamma
~
e^{ -\frac{d-2}{2d}
\sqrt{\frac{ d\Lambda }{
d-1
}}
t } ~
|k|, 
\nn
\Omega^\chi_k(L)
 &=&
 - \Gamma_\chi
 \left(2 (d-2) e^{\frac{1}{4}  \sqrt{\frac{d \Lambda }{d-1}} t}-d e^{ \sqrt{\frac{\Lambda }{(d-1) d}} t}\right) \times \nn &&
 e^{\frac{1}{4}  \sqrt{\frac{d \Lambda }{d-1}} t} 
 |k| (1-is) 
+ O(|k|^3), \nn
C_\Phi(t) &=& 
\frac{1}{2 A\sqrt{u^3 \Lambda}}
e^{\frac{3}{2} \sqrt{\frac{d \Lambda }{d-1}} t },  \nn
C_{\gamma_\perp}(t) &=&
\frac{1}{2 A (d+4) (d-1)^{3/2} d^{1/2} \Lambda^{1/2}}
e^{ \frac{3d+4}{2d} \sqrt{\frac{d \Lambda }{d-1}} t},
\label{eq:Omegat}
\nn
\eqa
where
$\Gamma=
\frac{A d^{1/2}}{2 (d+2) (d-1)^{1/2}  \Lambda^{1/2}}$,
$\Gamma_\chi =
 \frac{\sqrt{(d-2) (d-1)} 
 }{(d-4) \sqrt[4]{\frac{A^2 (d-1) d}{\Lambda }}}$
 and
$|k|=\sqrt{k_\mu k_\nu \bgimn(t)}$
with
$\bgimn(t)=\bgimn(L(t),L(t))$.
\eq{eq:Omegat} captures how the time-dependent energy of each mode changes adiabatically as the scale factor of the background geometry increases slowly with $t$.

\section{Physical consequences}

\begin{figure}[th]
    \centering
\includegraphics[width=0.8\linewidth]{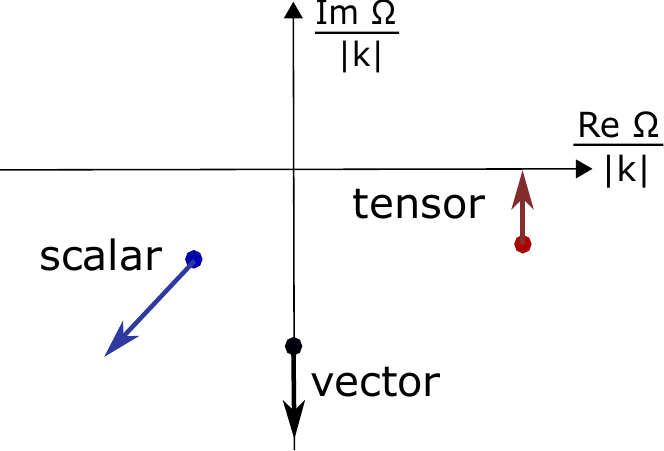} 
\caption{
The complex energies of the excitations for $s=-1$.
As time increases, 
the energies of the tensor graviton, the vector and the scalar shift toward 
the real axis,
along the imaginary axis
and 
at the $45^{\circ}$ in the lower half of the complex plane, respectively.
}
\label{fig:complexfrequency}
\end{figure}

All modes have imaginary components of energy.
The imaginary components are such that all modes exhibit damping over time, irrespective of $s$. 
Now, let us examine the spectrum of each mode and their physical consequences.
Different modes exhibit different levels of unitarity violation, as shown in  \fig{fig:complexfrequency}.
For the rest of this discussion, let us focus on the case with $s=-1$. 
All results discussed below can be straightforwardly extended to the case with $s=1$ through $i \rightarrow -i$.

The real part of the energy for the tensor gravitons is non-negative, as expected.
The tensor gravitons are excitations that are created close to the sub-Hilbert space ${\mathbb H}_0$ that satisfies ${\cal H}=0$ and ${\cal P}_\mu=0$, and are least affected by non-unitarity compared to other modes.
Consequently, the decay rate of the tensor graviton decreases exponentially in $t$,
and unitarity is restored in the large $t$ limit.
However, the non-zero damping at finite $t$ leads to observational consequences.
To see this, we note that the gravitational wave can be viewed as a coherent state of the tensor gravitons.
In the adiabatic limit, the traceless and transverse metric deformation satisfies the wave equation,
\bqa
\left[ \frac{\partial^2}{\partial t^2}
- c(t)^2 \bnabla^2
\right] h^{\mu}_{\nu}=0,
\eqa
where 
$c(t)
=1
- i  \Gamma
~
e^{ -\frac{d-2}{2d}
\sqrt{\frac{ d\Lambda }{
d-1
}}
t }
$ is the complex speed of the tensor graviton.
Due to the damping caused by the imaginary component,
the amplitude of the gravitational wave that 
was generated from a distant source at time $t_0$ and reaches us at time $t$ is suppressed as
\bqa
h^\mu_\nu(t)
=
F_k(t)
\tilde h^\mu_{\nu}(t),
\eqa
where $\tilde h^\mu_{\nu}(t)$ is the amplitude predicted by the unitary theory with $c=1$
and
$F_k(t)=e^{ \int_{t_0}^t \Im c(t') |k(t')| dt' }$
is the non-unitary correction, 
where
$|k(t')|=\sqrt{k_\mu k_\nu \bgimn(t')}$.
The suppression factor becomes
\bqa
F_k(t)=
e^{
-\frac{A 
|k(t)|
}{(d+2) \Lambda}
e^{ \sqrt{\frac{\Lambda}{d(d-1)}} t}
\left(
e^{ -\sqrt{\frac{d\Lambda}{4(d-1)}} t_0}
-
e^{ -\sqrt{\frac{d\Lambda}{4(d-1)}} t}
\right)
}.
\eqa
Its absolute value depends on the unknown parameter $A$,
and approaches $1$ in the small $A$ limit.
What is independent of $A$ is the relative suppression of the gravitational waves with different wavevectors. 
For the gravitational waves generated from the same source, the non-unitary corrections at wavevectors $k_1$ and $k_2$ are related by
\bqa
\frac{ \log F_{k_1}(t)}{ \log F_{k_2}(t)}
= 
\frac{|k_1|}{|k_2|}.
\eqa
The amplitude of the gravitational wave is predicted to be more strongly suppressed at high frequencies relative to low frequencies.

The other modes are excitations created mostly outside the sub-Hilbert space ${\mathbb H}_0$, and they are more strongly damped by non-unitary dynamics that cause the collapse of the wavefunction toward ${\mathbb H}_0$.
The energies of the longitudinal and transverse vector modes ($\Phi$ and $\gamma_\perp$) are purely imaginary, with their decay rates being
independent of wavevector but increasing exponentially in $t$.
Therefore, those modes decay toward 
$\pi_{\Phi}=0$
and
$\pi^\mu_{\gamma_\perp}=0$
uniformly at all length scales in space.
Because their decay rates are inversely proportional to $A$, we expect that the vector modes are practically unobservable for sufficiently small $A$.
Consequently, the violation of the momentum constraint decays to zero at the exponentially increasing rate.

\begin{figure}[th]
    \centering
\includegraphics[width=0.35\linewidth]{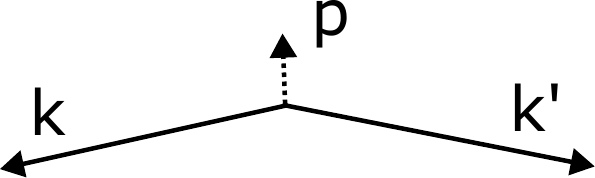} 
\caption{
The cubic vertex that produces a scalar of momentum $p$ with negative energy alongside two tensor gravitons of momenta $k$ and $k'$ with positive energies.}
\label{fig:graviton_production}
\end{figure}

The energy of the scalar mode ($\chi$) has a negative real part and an imaginary part.
They have the same magnitude and are proportional to the wavevector.
Usually, the presence of particles with negative energy is pathological because it can lead to the incessant creation of particles with positive energies (such as tensor gravitons) alongside particles with negative energy.
However, such processes are suppressed by the imaginary component of the energy.
To see this, let us consider the lowest order vertex that couples the tensor graviton and the scalar,
\bqa
H_{eff}' \sim \int dx e^{d \ba} \varphi 
h^\nu_\mu \bnabla^2 h^\mu_\nu.
\eqa
It can be obtained by expanding \eq{eq:tildeSalpha} to the cubic order in the fluctuation fields and using \eq{eq:tL}.
For simplicity, let us ignore the time-dependence in the mode frequencies and write the dispersions of the tensor graviton and the scalar as
\bqa
\Omega_k & = & |k|, \nn
\Omega^\chi_k & = & -v|k|(1+i), 
\eqa
where $v$ is the velocity of the negative energy mode at time $t$. 
We assume that $t$ is large enough that the decay rate of the tensor graviton can be ignored, and $v \gg 1$.
We have chosen the branch with $s=-1$, but the following conclusion remains unchanged for $s=1$.
To the lowest order, the rate at which the tensor graviton of momentum $k$ is produced between time $t$ and $t+T$ is given by
\bqa
\Gamma_k 
& \sim &
\frac{1}{T}
\int \frac{dk' dp}{(2\pi)^{2d}}
(k^2+k'^2)^2
\left|
\frac{
e^{-i (\Omega_k + \Omega_{k'} + \Omega^\chi_{p}) T} 
-1}
{\Omega_k + \Omega_{k'} + \Omega^\chi_{p}}
\right|^2 \times \nn &&
~~~~~~~~~~
(2\pi)^d \delta(k+k'+p). 
\label{eq:gammak}
\eqa
Here, 
the momenta of the second tensor graviton ($k'$) and the scalar ($p$) are summed over,
and the delta function imposes momentum conservation.
Due to the large velocity of the scalar mode, momentum conservation and the resonance condition 
($\Re (\Omega_k + \Omega_{k'} + \Omega^\chi_{p})=0$)
can be simultaneously satisfied for $k' \approx -k$ and $2|k| \approx v |p|$, as illustrated in \fig{fig:graviton_production}.
Without the imaginary component of the energy, 
the denominator of \eq{eq:gammak} would vanish at resonance, giving rise to a finite production rate.
However, the imaginary component of $\Omega_p^\chi$ smears the resonance.
Once the sum over $k'$ is removed with the delta function, the production rate can be written as
$
\Gamma_k \sim \frac{1}{T}
\int \frac{ d^dp}{(2\pi)^d} 
\frac{
( k^2 + ( k+p)^2 )^2
}{
(2|k| - v|p|)^2 + (v|p|)^2
}
$ at large $T$.
It is understood that there is a large momentum cutoff.
Importantly, the denominator does not vanish for any non-zero $k$: even if the resonance condition is satisfied at $v|p|=2|k|$ for the real part of the energy, the contribution from the imaginary component $(v|p|)^2$ remains non-zero.
As a result, the production rate $\Gamma_k$ approaches zero in the large $T$ limit.
This shows that the negative energy mode does not lead to instability, thanks to its finite lifetime.

The scalar mode, which represents a violation of the Hamiltonian constraint, acts as dark matter\cite{PhysRevD.109.124026,2025arXiv250504544I}.
Due to its relatively long lifetime at long wavelengths, it may contribute to the observed dark matter over long distance scales.
On the other hand, it can be difficult to detect them directly in laboratories with insufficient time resolution, because the lifetime of the mode at the laboratory scale is significantly shorter than that at astronomical scales.
For example, between the wavelengths of a few meters and a dark matter halo of size $100$ kpc, the ratio of the lifetimes is on the order of $10^{-21}$. 

Here, the large $d$ limit is considered to control our calculations.
In $d=3$, the $1/d$ corrections can be potentially significant.
In Appendix \ref{app:approx}, we identify the sources of such $1/d$ corrections.

\section{Summary and Discussion}

In this paper, we extend a recently proposed theory of time based on wavefunction collapse to general relativity.
It is based on the following postulations: 
(1) physical states generally violate the momentum and Hamiltonian constraints, and 
(2) time evolution is fundamentally a non-unitary process in which the violation of the constraints is gradually reduced through a wavefunction collapse induced by stochastic fluctuations of the lapse and shift functions.
Based on these postulations, it is shown that the scale factor of the universe monotonically increases under wavefunction collapse and acts as a clock.
The dynamics of the scalar, vector, and tensor graviton modes are described by a non-Hermitian Hamiltonian, with the arrow of time fixed by the initial state.
Unitarity emerges in the dynamics of the tensor gravitons as the imaginary component of their energy vanishes in the late time limit.
On the other hand, the vector and scalar modes, which represent constraint-violating excitations, show a stronger violation of unitarity.
While the energy of the vector mode is purely imaginary, the scalar mode exhibits marginally damped dynamics. 
The decay rate of the scalar mode is comparable to the real part of the energy and is proportional to the wavevector.
Due to the long lifetime at long wavelengths, the scalar excitation may have prominent observational effects at long-distance scales.

Here, we conclude with a few future directions.
Firstly, one can try to look for signatures of non-unitarity in the scattering amplitudes.
For example, the scattering amplitudes of the tensor graviton and the particles in the standard model will be affected by the short-lived, unstable particles.
At low energies, the effect is expected to be small due to the weak interaction; however, there should be observable effects at sufficiently high energies.
Secondly, it is of interest to extend the present study to configurations with strong spatial inhomogeneity.
While the principle remains the same, the inhomogeneity may affect the progression of wavefunction collapse in a space-dependent manner, especially near compact objects such as black holes.
Finally, one can consider more general cases in which the cosmological constant is not positive and the space has non-zero curvature.
Around a homogeneous and isotropic space, the curvature scalar can be written as
$R = K e^{-2\alpha} + \tilde R$, 
where the first term originates from the curvature of the background space, and $\tilde R$ is the contribution of fluctuations.
Then, the Hamiltonian density is written as
\bqa
{\cal H} =
u
\Bigl[
\tilde \Lambda(\alpha)
+
e^{-2d\alpha}
\Gmnrs
\td \Pi^{\mu \nu} 
\td \Pi^{\rho \sigma} 
-  \tilde R 
\Bigr]-
e^{-2d\alpha}
\Pi^2, \nn
\eqa
where 
$\tilde \Lambda(\alpha) 
=
\Lambda - K e^{-2\alpha}$ is the effective vacuum energy. 
For the saddle-point without excitation,
we can set $\Gmnrs
\td \Pi^{\mu \nu} 
\td \Pi^{\rho \sigma} 
-  \tilde R 
=0$.
Even for $\Lambda<0$, 
$\tilde \Lambda(\alpha)$ needs to be non-negative;
if $\tilde \Lambda(\alpha) < 0$, there is no  configuration that satisfies ${\cal H}=0$ 
and the wavefunction collapse results in $\Psi=0$ in the late time limit.
If an initial state satisfies $\tilde \Lambda(\alpha) > 0$ with a sufficiently negative $K$,
the evolution generated by the wavefunction collapse is similar to the case with 
a positive cosmological constant.
In particular, $\alpha$ still increases under the wavefunction collapse, 
and the direction of time evolution is determined by the initial state in the same way.
However, 
the dependence of $\ba$ on $L$ will be modified\cite{glfz-yvnl} because $\tilde \Lambda(\alpha)$ decreases with increasing $\alpha$. 
More importantly, a qualitatively new behavior is expected to arise as the universe approaches the critical size $\alpha_c$ at which $\tilde \Lambda (\alpha_c)=0$. 
Near $\alpha_c$, the higher order terms in $\partial_l \alpha/\tilde \Lambda(\alpha)$ play important roles in determining the dynamics of $\alpha$ in \eq{eq:saddlepts}.
The same issue arises in flat space with $\Lambda=0$.
It would be of great interest to understand how the late time dynamics differ in the spaces with $\Lambda \leq 0$ compared to the one with $\Lambda >0$.
%

%
%

\section*{Acknowledgments}

The research was supported by
the Natural Sciences and Engineering Research Council of
Canada.
Research at Perimeter Institute is supported in part by the Government of Canada through the Department of Innovation, Science and Economic Development Canada and by the Province of Ontario through the Ministry of Colleges and Universities.

\section*{DATA AVAILABILITY}

No data were created or analyzed in this study.

\bibliographystyle{unsrtnat}

\bibliography{references_high}

\appendix
\section{Normal ordering of ${\cal H}$}
\label{app:1}

To write \eq{eq:Psiepsilons0} in the path-integral representation,
it is convenient to use the normal ordering in \eq{eq:He2}.
This is because the space of positive definite spatial metrics has singular boundaries.
To see the complications that arise from different normal orderings, let us consider an alternative representation of \eq{eq:He2}:
$\hat H[\epsilon] = 
\int d^dx~  \epsilon(x) 
\left[ \hPimn \hPirs \frac{ \hGmnrs }{\hsg}  
+
ik^2 A_0
\hat \Pi \frac{1}{\sg}
+ ..
\right]
$,
where 
$\hPimn$'s are placed to the left of $\hgmn$'s.
In this representation,
the coefficient of 
$\hat \Pi \frac{1}{\sg}$
is reversed.
Now, let us consider the first term,
 $\langle g^{(1)} | e^{-i \hat H[\epsilon^{(1)}] } |\Psi_0 \rangle$
 that appears in the path-integral representation.
In the small $\epsilon^{(1)}$ limit, we need to evaluate
\bqa
 && \langle g^{(1)} | \hat H[\epsilon^{(1)}]  |\Psi_0 \rangle
=
-   \int d^dx~  \epsilon^{(1)}(x)  
\int 
D g^{(0)}
D \Pi^{(1)} \nn &&
e^{
i \Pi^{(1)\mu \nu}
(g^{(1)}_{\mu \nu}-
g^{(0)}_{\mu \nu} )
}
\frac{\delta}{\delta g^{(0)}_{\mu\nu}}
\frac{\delta}{\delta g^{(0)}_{\rho\sigma}}
\left[
\frac{ \Gmnrs^{(0)}  }{\sqrt{g^{(0)}}} 
\Psi_0(g^{(0)})
\right]+.., \nn
\label{eq:firststep}
\eqa
where the delta functional 
$
\delta \left( 
g^{(1)}_{\mu \nu}-
g^{(0)}_{\mu \nu} \right)
=
\int 
D \Pi^{(1)}
e^{
i \Pi^{(1)\mu \nu}
(g^{(1)}_{\mu \nu}-
g^{(0)}_{\mu \nu} )
}
$ is inserted.
To convert
$-i\frac{\delta}{\delta g^{(0)}_{\mu\nu}}$ to
$\Pi^{(1)\mu\nu}$,
one has to perform
the integration by part,
which generates a total derivative term, 
$\int Dg^{(0)}
\frac{\delta}{\delta g^{(0)}_{\mu\nu}}
\left[
e^{
-i \Pi^{(1)\mu \nu}
g^{(0)}_{\mu \nu} 
}
\frac{\delta}{\delta g^{(0)}_{\rho\sigma}}
\frac{ \Gmnrs^{(0)}  }{\sqrt{g^{(0)}}} 
\Psi_0(g^{(0)})
\right]
$.
The problem is that the boundary contribution is non-negligible for general normalizable wavefunctions because 
$\frac{1}{\sqrt{g^{(0)}}}$ 
 become singular in the small $g^{(0)}$ limit.
The representation in \eq{eq:He2} has no such issue.
The same procedure applied to \eq{eq:He2} leads to
\bqa
&&  \langle g^{(1)} | \hat H[\epsilon^{(1)}]  |\Psi_0 \rangle
=
-  
\int d^dx~  \epsilon^{(1)}(x)  
\frac{ \Gmnrs^{(1)}  }{\sqrt{g^{(1)}}}
\int 
D g^{(0)}
D \Pi^{(1)} 
\nn &&
e^{ i \Pi^{(1)\mu \nu} (g^{(1)}_{\mu \nu}- g^{(0)}_{\mu \nu} ) }
\left[
\frac{\delta}{\delta g^{(0)}_{\mu\nu}}
\frac{\delta}{\delta g^{(0)}_{\rho\sigma}}
\Psi_0(g^{(0)})
\right]+...
\label{eq:g1Hpsi0}
\eqa
Unlike
\eq{eq:firststep},
one can perform the integration by part for square integrable $\Psi_0$ to obtain
\eq{eq:pathintegral}.

\section{Effective Hamiltonian}
\label{app:Heff}


In this appendix, we derive \eq{eq:Heff} from 
\eq{eq:PsiL3V}.
This can be most easily done by rewriting \eq{eq:PsiL3V} in the Hamiltonian picture as
\bqa
|\tilde  \Psi_s(L) \rangle 
=
\bar w_s(L)
\hat W
e^{
is
\int_0^L dl 
~ \hat H_2(l;L)
}
| \tilde \Psi'_{0,s} \rangle.
\label{eq:HforlL}
\eqa
Here, 
$w_s(L)=
e^{ 
i \int_0^{L} dl \int dx 
\xmd
\bar {\cal L} 
}$.
%
The quadratic `Hamiltonian' is
\bqa
&& \hat H_2(l;L)
=  
-s
\int dx  e^{d \ba} 
\left\{ \EL
e^{-2d \ba}
\hat \pi_{\mu}^{~\nu} \hat \pi_{\nu}^{~\mu} 
- \frac{\EL}{4} \hat h_{\mu}^{~\nu} \bnabla^2 \hat h_{\nu}^{~\mu} \right. \nn &&
-i 
e^{-2d \ba}
 \hat \pi_{ \chi} 
\frac{-\bnabla^2+4d^2u\Lambda}  {4 d^2}
 \hat \pi_{ \chi} 
- \EL    (d-1)(d-2)   
 (\bnabla \hat{\chi})^2
 \nn &&
\left. -i e^{-2d \ba} \left[
\frac{1}{4}
 (\hat \pi_{\Phi})^2 
+ \bgimn
 \hat \pi_{\gamma_\perp \mu}
 \hat \pi_{\gamma_\perp \nu}
 \right] \right\}.
\label{eq:H2}
\eqa
$\hat \pi_{\mu}^{~\nu}$,
$\hat \pi_\chi$,
$\hat \pi_\Phi$ and
$\hat \pi_{\gamma_\perp \mu}$
are the conjugate momenta of
$\hat h_{\nu}^{~\mu}$,
$\hat \chi$, 
$\hat \Phi$
and $\hat \gamma_\perp^{\mu}$,
respectively.
$\hat H_2(l;L)$ depends on both $l$ and $L$ through $\ba(l;L)$.
One may consider a state realized at an intermediate step of evaluating \eq{eq:HforlL},
such as
$
\bar w_s(L')
\hat W
e^{
is
\int_0^{L'} dl 
~ \hat H_2(l;L)
}
| \tilde \Psi'_{0,s} \rangle
$ for $0< L'<L$.
However, this is not the same as the state $|\Psi_s(L')\rangle$ realized at time $L'$, and does not have physical significance.
This also implies that $\hat H_2(l;L)$ does not represent the Hamiltonian for the actual time evolution.

$\hat H_2(l;L)$ is already diagonalized for $\Phi$ and $\gamma_\perp$.
To diagonalize it for the tensor graviton and the scalar, we first go to Fourier space,
\bqa
\hat h_{\nu}^{~\mu}(x)
&= &
\frac{1}{V} 
\sum_{j=1}^{
(d-2)(d+1)/2}
\sum_{k}
~e^{i k x} 
\epsilon^{(j) \mu}_{k,\nu}
\hat h^{(j)}_k, \nn
\hat \chi(x)
&=& \frac{1}{V} 
\sum_k
~e^{i k x} 
\hat \chi_{k}.
\eqa
Here, we assume that the space has the topology of the $d$-dimensional torus with coordinate volume $V$.
The momentum $k$ takes discrete values: 
$k= \frac{2\pi}{V^{1/d}}(n_1,..,n_d)$ with
$n_i \in \mathbb{Z}$
and $n_i \sim n_i + V^{1/d}$.
$\epsilon^{(j) \mu}_{k, \nu}$ is the 
polarization tensor for the $j$-th traceless and transverse gravitational mode.
In the momentum space, the Hamiltonian for the tensor and scalar $\chi$ reads
\bqa
&&\hat H_2(l;L)
= \nn
&&
-\frac{s}{V}
\sum_k
\Bigg\{
\sum_j
\EL
\left[
e^{-d \ba}
\hat \pi_{k}^{(j)} 
\hat \pi_{-k}^{(j)} 
+ 
e^{d \ba}
\frac{k^2}{4} 
\hat h_{k}^{(j)} 
\hat h_{-k}^{(j)} 
\right] 
\nn && 
~~~~~~~~~~
~~~~~~~~~~
-i \frac{
k^2
+4d^2u\Lambda}  {4 d^2}
e^{-d \ba}
 \hat \pi_{\chi,k}
 \hat \pi_{\chi,-k}
\nn && 
~~~~~~~~~~
~~~~~~~~~~
- e^{d \ba}
 \EL   
 (d-1)(d-2) k^2
 \hat \chi_{k} 
 \hat \chi_{-k} 
 \Bigg\} \nn
&& + i s
\int dx  e^{-d \ba} 
\left\{
\frac{1}{4}
 (\hat \pi_{\Phi})^2 
+ \bgimn
 \hat \pi_{\gamma_\perp \mu}
 \hat \pi_{\gamma_\perp \nu}
  \right\},
\label{eq:H3}
\eqa
where 
$k^2 \equiv 
\bar g^{\mu \nu}(l;L) k_\mu k_\nu
$.
The quadratic Hamiltonian can then be diagonalized as
\bqa
\hat H_2(l;L)
&=&
\sum_k
\Bigg[
\sum_j 
\omega_{k}(l;L)
\left(
\hat a^{(j) \dagger}_{k}(l;L)
\hat a^{(j)}_{k}(l;L)
+ \frac{1}{2}
\right) 
\nn &&
~~~~~~~~~~~~~
+
\omega^\chi_k(l;L)
\left(
\hat b_k^{\dagger}(l;L)
\hat b_k(l;L)
+ \frac{1}{2}
\right)
\Bigg] \nn 
&& 
+is 
\int dx  e^{-d \ba} 
\left\{
\frac{1}{4}
 (\hat \pi_{\Phi})^2 
+ \bgimn
 \hat \pi_{\gamma_\perp \mu}
 \hat \pi_{\gamma_\perp \nu}
  \right\}.
\label{eq:Hl}
\eqa
Here,
\bqa
\omega_{k}(l;L)
&=&
\left[
\sqrt{\frac{u}{\Lambda}} 
(\partial_l \ba)
- i s
\frac{(\partial_l \ba)^2}{4 u \Lambda^2} \right] |k(l;L)|,  \nn
\omega^\chi_k(l;L) &= &
-\sqrt{2(d-1)(d-2)}
u^{3/4}
\Lambda^{1/4}
(\partial_l \ba)^{1/2} 
\times \nn &&
\sqrt{\left( 1+\frac{|k(l;L)|^2}{4d^2u\Lambda} \right)}
|k(l;L)|
(1 - is)
\label{eq:omegal}
\eqa
represent the complex frequencies
with
$|k(l;L)| \equiv \sqrt{ \bar g^{\mu \nu}(l;L) k_\mu k_\nu }$.
$\hat a^{(j)}_{k}(l;L)$ and $
\hat b_k(l;L)$
($
\hat a^{(j) \dagger}_{k}(l;L)$ and $
\hat b_k^\dagger(l;L)$)
are annihilation (creation) operators of the tensor graviton of polarization $j$ and scalar mode, respectively,
\bqa
\hat a^{(j)}_{k}(l;L)
&=&
\frac{1}{\sqrt{2V}}
\left(
\sqrt{f_{k}} \hat h^{(j)}_k
+ \frac{i}{\sqrt{f_{k}}} \hat \pi^{(j)}_{k}
\right),   \nn
\hat a^{(j) \dagger}_{k}(l;L)
&=&
\frac{1}{\sqrt{2V}}
\left(
\sqrt{f_{k}} \hat h^{(j)}_{-k}
- \frac{i}{\sqrt{f_{k}}} \hat \pi^{(j)}_{-k}
\right),  
\nn
\hat b_k(l;L)
&=&
\frac{1}{\sqrt{2V}}
\left(
\sqrt{f_{\chi,k}} \hat \chi_{k}
+ \frac{i}{\sqrt{f_{\chi,k}}} \hat \pi_{\chi,k}
\right), \nn
\hat b_k^\dagger(l;L)
&=&
\frac{1}{\sqrt{2V}}
\left(
\sqrt{f_{\chi,k}} \hat \chi_{-k}
- \frac{i}{\sqrt{f_{\chi,k}}} \hat \pi_{\chi,-k}
\right),
\eqa
where 
$f_{k}=
\frac{e^{d\ba}|k(l;L)|}{2}$
and
$f_{\chi,k}=
\sqrt{i s}
\sqrt{\frac{
 (d-1)(d-2) |\EL| }{u \Lambda 
 \left( 1+\frac{|k(l;L)|^2}{4d^2u\Lambda} \right)}}
e^{d\ba}
|k(l;L)|
$.
In $f_{\chi, k}$, the sign ambiguity in $\sqrt{is}$ should be resolved such that 
the vacuum wavefunction 
$
\Psi_{vac}(\chi) \sim 
e^{-\frac{1}{2}\sum_k f_{\chi,k} \chi_k \chi_{-k}}
$ annihilated by 
$\hat b_k$
is normalizable.
This forces the real part of $f_{\chi, k}$ to be positive and 
$\sqrt{i s} = e^{i s \pi/4} = \frac{1}{\sqrt{2}}(1+ is)$.
While 
$\hat a^{(j)\dagger}_{k}$
is the adjoint of
$\hat a^{(j)}_{k}$,
$\hat b_k^\dagger$ is not the adjoint of $\hat b_k$ because $f_{\chi,k}$ is not real.
This is because the Hamiltonian that governs the scalar is not a multiple of a Hermitian operator.

All frequencies in \eq{eq:omegal} have imaginary components.
This is expected because the dynamics are induced by the non-unitary wavefunction collapse.
The real part of the tensor graviton frequency is positive, but the imaginary component has the sign such that the mode grows exponentially under 
\eq{eq:HforlL}.
Furthermore, the scalar mode has a negative real frequency.
This is due to the fact that the scalar has the wrong sign in the spatial gradient term in \eq{eq:H2}.
If $\hat H_2(l;L)$ were the actual Hamiltonian for time evolution, these would imply the existence of tachyonic modes.
However, we cannot reach a conclusion about the stability of the theory based on \eq{eq:HforlL} because it does not describe the time evolution.
It is merely an intermediate step in calculating $|\tilde \Psi_s(L) \rangle$.
The physical time evolution is $|\tilde  \Psi_s(L) \rangle$ as a function of time $L$.
In particular, the path of the physical time evolution, which is represented by the solid line in  \fig{fig:Gamma_smallL}, is not the same as the paths followed in \eq{eq:HforlL}, which is denoted by 
the dashed lines in \fig{fig:Gamma_smallL}. 
For the same reason, \eq{eq:omegal} does not represent the physical energies of the excitations.
We will later see that there exists no instability under the actual time evolution.

From Eqs. (\ref{eq:HforlL}) and (\ref{eq:H2}),
the effective Hamiltonian 
defined in \eq{eq:Heff0}
with respect to time $t$
in \eq{eq:tL} 
can be written as
\bqa
&&\hat H_{eff}(t)
= 
\hat W
\Bigg\{
\hat H_2(L;L) 
+
\int_0^L dl ~
\nn &&
\left.
\hat U(L,l) 
\left[
\partial_L \hat H_2(l;L)
\right]
\hat U^{-1}(L,l)
\Bigg\}
\left( \frac{dL}{dt} \right) 
\hat W^{-1} \right|_{L=L(t)}. \nn
\label{eq:fullHeff}
\eqa
Here,
$
\hat U(l_2;l_1) 
=
{\cal T}_l\left[
e^{-i \int_{l_1}^{l_2} dl'
\hat H_2(l';L)
} \right]
$.
The operators in ${\cal T}_l\left[..\right]$ are ordered such that
$\hat H_2(l;L)$ with larger $l$ are placed on the left of those with smaller $l$.
In general, \eq{eq:fullHeff} includes terms that describe 
the pair creation/annihilation of particles caused by the $l$-dependent background.
In the large $d$ limit, however,
$\omega_{k}(l;L), 
\omega^\chi_k(l;L)
\gg 
\partial_l \ba$, 
and one can use the adiabatic approximation. 
In the adiabatic limit, the pair-creation and annihilation terms are negligible, and one obtains
\bqa
&& \hat H_{eff}(t) 
=
\sum_k
\Bigg[
\sum_j 
\Omega_{k}(t)
\left(
\hat a^{(j) \dagger}_{k}(t)
\hat a^{(j)}_{k}(t)
+ \frac{1}{2}
\right) 
\nn &&
~~~~~~~~~~ 
~~~~~~~~~~ 
~~~~~~~~~~ 
+
\Omega^\chi_k(t)
\left(
\hat b'_k{}^{\dagger}(t)
\hat b'_k(t)
+ \frac{1}{2}
\right)
\Bigg] \nn &&
+is \int dx \frac{1}{\sqrt{\bar g(t)}}
\left[
C_\Phi(t)
 (\hat \pi'_{\Phi})^2 
+ 
C_{\gamma_\perp}(t)
\bgimn(t)
 \hat \pi_{\gamma_\perp \mu}
 \hat \pi_{\gamma_\perp \nu}
 \right]. \nn
\label{eq:Heffapp}
\eqa
Here,
$ \hat a^{(j)}_{k}(t) \equiv  \hat a^{(j)}_{k}(L(t),L(t))$,
$ \hat b'_{k}(t) \equiv  \hat W \hat b_{k}(L(t),L(t)) \hat W^{-1}$ represent the annihilation operators of the physical modes at time $t$.
 $\hat \pi'_{\Phi} 
 = \hat W 
 \hat \pi_{\Phi}  \hat W^{-1}$.
$\bgmn(t)=\bgmn(L(t),L(t))$.
The mode energies are written as
\bqa
\Omega_{k}(t) &=&
\dot L \left[
 \frac{d}{dL}
 \int_0^{L} dl ~
 \omega_{k}(l;L) 
 \right]_{L=L(t)}, \nn
\Omega^\chi_k(t)
 &=& 
\dot L \left[
 \frac{d}{dL}
 \int_0^{L} dl ~
 \omega^\chi_{k}(l;L) \right]_{L=L(t)}, \nn
C_\Phi(t) &=& 
e^{d \ba(L;L)}
\dot L \left[ 
\frac{d}{dL}
\int_0^{L} dl ~
e^{-d \ba(l;L)}
\right]_{L=L(t)}, \nn
C_{\gamma_\perp}(t) &=&
e^{(d+2) \ba(L;L)}
\dot L \left[
\frac{d}{dL}
\int_0^{L} dl ~
e^{-(d+2) \ba(l;L)}
\right]_{L=L(t)}, 
\nn
\label{eq:Omegak}
\eqa
where $\dot L = \frac{dL(t)}{dt}$.  
In the large $t$ limit, the mode frequencies are given by
\eq{eq:Omegat}.

The energy of each mode in \eq{eq:Omegak} has two contributions.
The first is the `boundary' contribution from $l=L(t)$.
The real part of the tensor graviton energy is entirely generated from this boundary term.
The second is the `bulk' contribution that arises from $0<l<L(t)$.
The imaginary part of the tensor graviton energy and the energy of all other modes are dominated by this second contribution. 
The bulk term arises because $\ba(l;L)$ at a fixed $l$ depends on $L$ as is shown in \fig{fig:Gamma_smallL}.
Since the net decay of the wavefunction at $L$ depends on the entire path $\ba(l;L)$ as a function of $l$, the decay rate depends on $L$ through the $L$-dependent saddle-point path.
In a sense, the decay rate is determined through a non-Markovian way as it depends on the entire saddle-point history of $\ba(l;L)$.
However, we emphasize that $\ba(l;L)$ as a function of $l$ does not represent the actual time evolution because the latter is given by $\ba(L(t);L(t))$ as a function of $t$.
For the actual time evolution, $|\Psi_s(t+dt) \rangle$ is entirely determined from $|\Psi_s(t) \rangle$. 
%
%
The $t$-dependent decay rates are powers of 
$e^{\sqrt{\frac{d \Lambda }{d-1}}t}$
because they depend on $L$ algebraically, and 
$L(t) \sim  e^{\sqrt{\frac{d \Lambda }{d-1}}t} $.

The fact that $\Im \Omega_k$ 
in \eq{eq:Omegat}
and $\Im \omega_k$ in
\eq{eq:omegal}
for the tensor graviton have opposite signs can be understood as follows.
As $L$ increases, $\ba(l;L)$ decreases at fixed $l$, which, in turn, causes $\partial_l \ba$ and the imaginary component of 
\eq{eq:omegal} to decrease.
As a result, $\Im \Omega_k$ in \eq{eq:Omegak} has the sign opposite to $\Im \omega_k$.

\section{$1/d$ corrections}
\label{app:approx}

In this appendix, we revisit the approximations employed in the large $d$ limit and discuss how $1/d$ corrections arise at finite $d$.

\begin{itemize}
\item Gradient expansion  of the saddle point;

In \eq{eq:saddlepts},
the saddle-point solution for $\Pi$ only keeps the term to the linear order in 
$\partial_l \ba/\Pic^3$.
This approximation is justified if
\bqa
\partial_l \ba 
\ll (u \Lambda)^{3/2},
\label{eq:approx1}
\eqa
where $u=d(d-1)$.
According to \eq{eq:balL},
\bqa
\partial_l \ba \leq 
\frac{\left( \sqrt{ 1+ 4A u d \Lambda L e^{ -d \alpha_0} } -1 \right)}{dL},
\label{eq:dalphabound}
\eqa
and
\eq{eq:approx1} is guaranteed to be satisfied in the large $d$ limit.
For a finite $d$ with
$\sqrt{u \Lambda} < 
2 A e^{ -d \alpha_0}$,
\eq{eq:approx1} is violated 
at small $L$.
In that case, 
the higher order terms in 
$\partial_l \alpha/\Pic^3$ 
become important for the `early' universe with 
$L \leq 
\frac{2 e^{-d \alpha_0} 
\left( 
2 A- e^{d \alpha_0} \sqrt{\Lambda  u}
\right)
}{d \Lambda ^2 u^2}$.

\item 
Adiabatic approximation;

\eq{eq:Heff} is obtained based on the adiabatic approximation, 
which is valid for the modes whose frequencies in $l$ 
are much greater than that rate at which $\ba$ changes,
\bqa
\partial_l \ba
\ll
\omega_{k}(l;L),
\omega^\chi_k(l;L),
\label{eq:approx2}
\eqa
where
$\omega_{k}(l;L)$
and
$\omega^\chi_k(l;L)$
are given in \eq{eq:omegal}.
\eq{eq:approx2} is satisfied for the modes with
$ |k(l;L)| 
\gg
\sqrt{\Lambda/u},
~
(\partial \ba)^{1/2}
\Lambda^{-1/4}
u^{-5/4} $,
where 
$|k(l;L)| 
=
\sqrt{ \bar g^{\mu \nu}(l;L) k_\mu k_\nu }$.
In the large $d$ limit, the adiabatic approximation is valid for all non-zero $k$ both for the tensor and scalar gravitons.
For a finite $d$, one should include non-adiabatic processes in which the expanding geometry creates pairs of particles at low momenta below crossover momentum scales.
For the tensor graviton, the crossover momentum is fixed to be $\sqrt{\Lambda/u}$.
For the scalar graviton, the crossover momentum increases with decreasing $l$ as the universe expands faster at earlier times.
Therefore, the non-adiabatic corrections are present over a larger range of momenta at smaller $l$.
However, the crossover momentum is still bounded from above because 
\eq{eq:dalphabound} is bounded.

\item 
Quadratic approximation;

\eq{eq:Heff} has been derived  from the quadratic action.
In general, the cubic and higher-order interaction vertices in \eq{eq:Salpha0} renormalize the effective action, including the kinetic energy.
Since the vector graviton is non-propagating at long distance scales, we focus on the scalar and tensor gravitons.
The Lagrangian density that includes the cubic and quartic vertices for the propagating modes can be written as
\bqa
{\cal L}' &\sim &
\delta \pi \partial_l \varphi
+ 
\pi_\perp^{\mu \nu}  \partial_l 
h_{ \mu \nu}
- \EL \left[  \Ht +  D_4 \Ht^2 \right]
\nn &&
+ i \left[ 
C_2
\delta \pi^2
+
C_3
\delta \pi^3
+
C_4
\delta \pi^4
\right], \nn
\Ht & \sim &
\pi_{\perp,\nu}^{\mu}
\pi_{\perp,\mu}^{\nu}
- \frac{1}{4}
 h_{ \nu}^{\mu}
 \bnabla^2
 h_{ \mu}^{\nu} 
-B_2\left| \bnabla \da \right|^2,
\label{eq:Sint}
\eqa
where
$\delta \pi =  
e^{-d\ba} \delta \Pi$,
$\pi_\perp^{\mu \nu}  
=
e^{-d\ba} \Pi_\perp^{\mu \nu}$,
$C_2 \sim u \Lambda$,
$C_3 \sim \sqrt{u \Lambda}$,
$C_4 \sim 1$,
$B_2 \sim (d-1)(d-2)$,
$D_4 \sim \Lambda^{-1}$
and  $\EL$ is defined in 
\eq{eq:EL}.
The non-quadratic terms inside $\Ht$ are dropped because they are suppressed below the Planck energy scale $l_p^{-1}$, which is set to be $1$ here.
At low energies, the interactions from $\Ht^2$ are more important than those.

In the adiabatic limit, 
the scaling dimensions of the fields are given by
$ [\varphi]= [h^\mu_\nu]= (d-1)/2$ and $[\delta \pi] = [\pi^\mu_{\perp,\nu}] = (d+1)/2$.
Under the scale transformation that leaves the quadratic terms invariant,
$C_3, C_4, D_4$ carry negative scaling dimensions, and
the quantum corrections from them become negligible below crossover energy scales set by the dimensionful coefficients. 
The self-energy for the scalar 
mode generated from the non-quadratic terms of $\delta \pi$ becomes negligible compared with the bare kinetic energy for
$|k(l;L)| \ll
\left[
u^{\frac{3}{4}}
\Lambda^{\frac{7}{4}}
\partial_l \ba^{-\frac{1}{2}}
\right]^{\frac{1}{d+1}}
$.
Here, the factor of $\partial_l \ba^{-\frac{1}{2}}$ arises due to the fact that 
the momentum scale is related to the frequency scale through $k  \sim 
(\partial_l \ba)^{-\frac{1}{2}}
\omega$ 
in \eq{eq:omegal}
for the scalar mode.
%
On the other hand, the self-energy from $\Ht^2$ is suppressed for $|k(l;L)|  \ll 
\Lambda^{\frac{1}{d+1}}$.
In the large $d$ limit, the crossover energy scale becomes large and approaches the Planck scale. 
For finite $d$ and small $\Lambda$, however, the crossover energy scales
become significantly lower than the Planck scale;
for $d=3$ and
$\Lambda \sim 10^{-120}$,
the self-energy of the graviton becomes important above  $10^{-30}$ 
in the Planck unit,
which becomes
$\sim 10^{-2} eV$ for $l_p^{-1} \sim 10^{28} eV$.
This implies that the gravitons become incoherent above the crossover energy scale due to interactions.

\end{itemize}

\end{document}